\documentclass[sigconf, nonacm]{acmart}
\usepackage[T1]{fontenc}
\usepackage{listings}
\usepackage{subcaption}
\usepackage{pgfplots}
\usepackage{booktabs}
\usepackage{amsmath}
\usepackage{float} 
\usepackage[export]{adjustbox}
\usepackage{lipsum}
\usepackage{xcolor}
\usepackage{multirow}
\usepackage{overpic}

\newcommand{\mult}{{\ooalign{$\phantom{0}$\cr\hidewidth$\scriptstyle\times$\cr}}}
\AtBeginDocument{
  \providecommand\BibTeX{{
    \normalfont B\kern-0.5em{\scshape i\kern-0.25em b}\kern-0.8em\TeX}}}

\begin{document}
    \title[RT-Percept, Arxiv Version]{Training and Predicting Visual Error for Real-Time Applications}
    \author{\href{http://www.jaliborc.com/}{Jo\~{a}o Lib\'{o}rio Cardoso}}
\orcid{0000-0002-6530-7244}
\affiliation{
  \institution{TU Wien}
  \country{Austria}
}

\author{Bernhard Kerbl}
\orcid{0000-0002-5168-8648}
\affiliation{
  \institution{TU Wien}
  \country{Austria}
}

\author{Lei Yang}
\orcid{0000-0002-2523-671X}
\affiliation{
  \institution{NVIDIA Corp}
  \country{USA}
}

\author{Yury Uralsky}
\orcid{0000-0001-7142-6998}
\affiliation{
  \institution{NVIDIA Corp}
  \country{USA}
}

\author{\href{https://www.cg.tuwien.ac.at/staff/MichaelWimmer}{Michael Wimmer}}
\orcid{0000-0002-9370-2663}
\affiliation{
  \institution{TU Wien}
  \country{Austria}
}

\renewcommand{\shortauthors}{Lib\'{o}rio Cardoso et al.}
\begin{abstract}
Visual error metrics play a fundamental role in the quantification of perceived image similarity. Most recently, use cases for them in real-time applications have emerged, such as content-adaptive shading and shading reuse to increase performance and improve efficiency. A wide range of different metrics has been established, with the most sophisticated being capable of capturing the perceptual characteristics of the human visual system. However, their complexity, computational expense, and reliance on reference images to compare against prevent their generalized use in real-time, restricting such applications to using only the simplest available metrics. In this work, we explore the abilities of convolutional neural networks to predict a variety of visual metrics without requiring either reference or rendered images. Specifically, we train and deploy a neural network to estimate the visual error resulting from reusing shading or using reduced shading rates. The resulting models account for 70\%--90\% of the variance while achieving up to an order of magnitude faster computation times. Our solution combines image-space information that is readily available in most state-of-the-art deferred shading pipelines with reprojection from previous frames to enable an adequate estimate of visual errors, even in previously unseen regions. We describe a suitable convolutional network architecture and considerations for data preparation for training. We demonstrate the capability of our network to predict complex error metrics at interactive rates in a real-time application that implements content-adaptive shading in a deferred pipeline. Depending on the portion of unseen image regions, our approach can achieve up to $2\times$ performance compared to state-of-the-art methods.
\end{abstract}


\begin{teaserfigure}
  \centering
  \begin{subfigure}{0.33\textwidth}
    \centering
    \includegraphics[width=\textwidth]{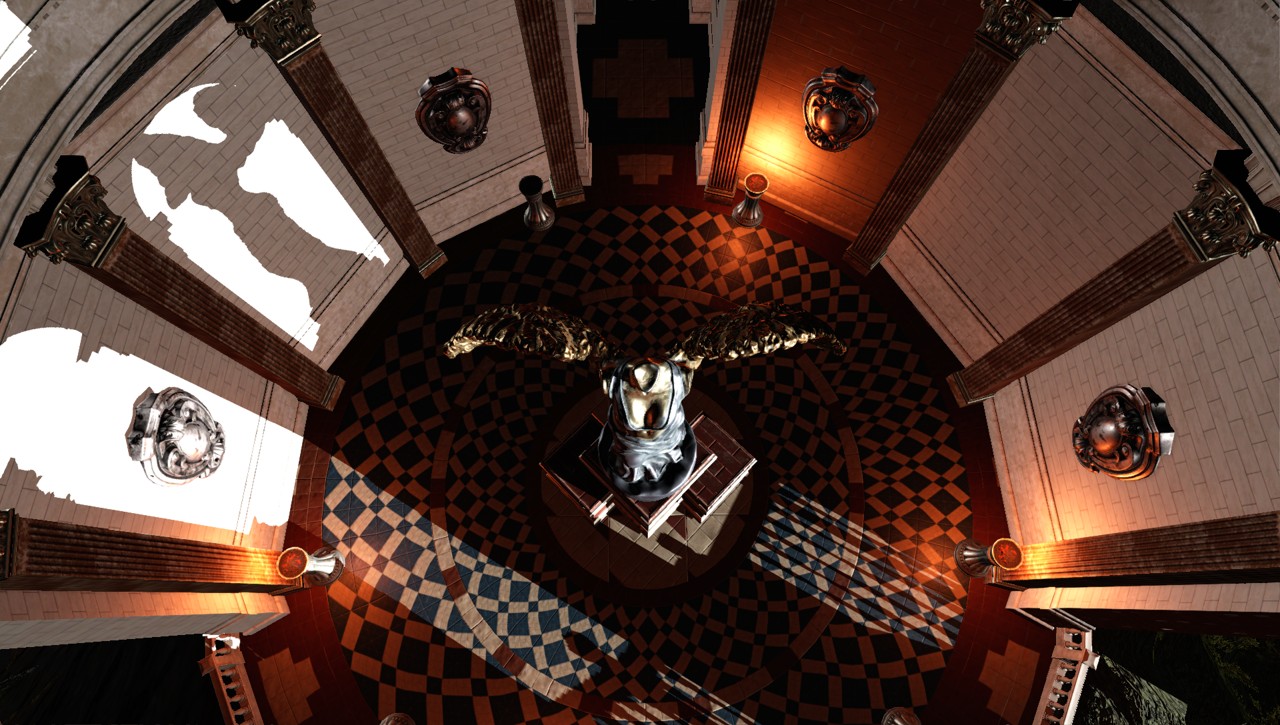}
    \caption{Suntemple scene}
  \end{subfigure}
  \begin{subfigure}{0.33\textwidth}
    \centering
    \includegraphics[width=\textwidth]{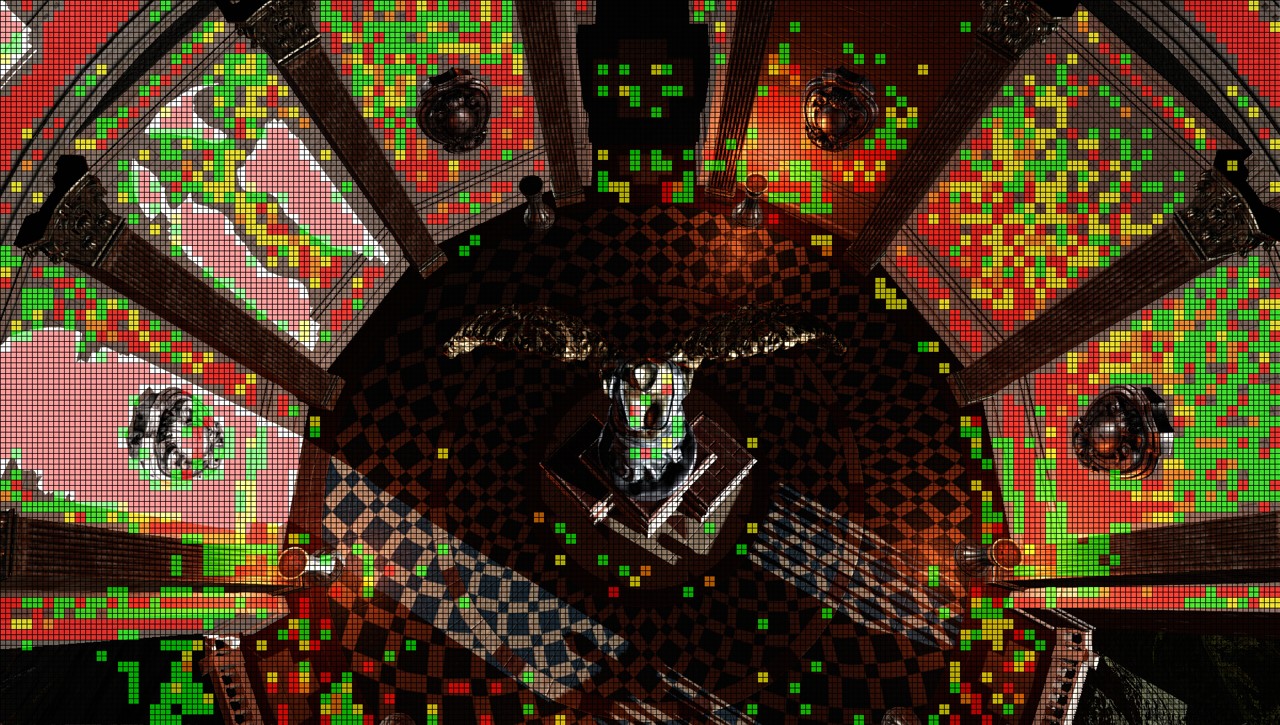}
    \caption{VRS with JNFLIP (predicted)}
  \end{subfigure}
  \begin{subfigure}{0.33\textwidth}
    \centering
    \includegraphics[width=\textwidth]{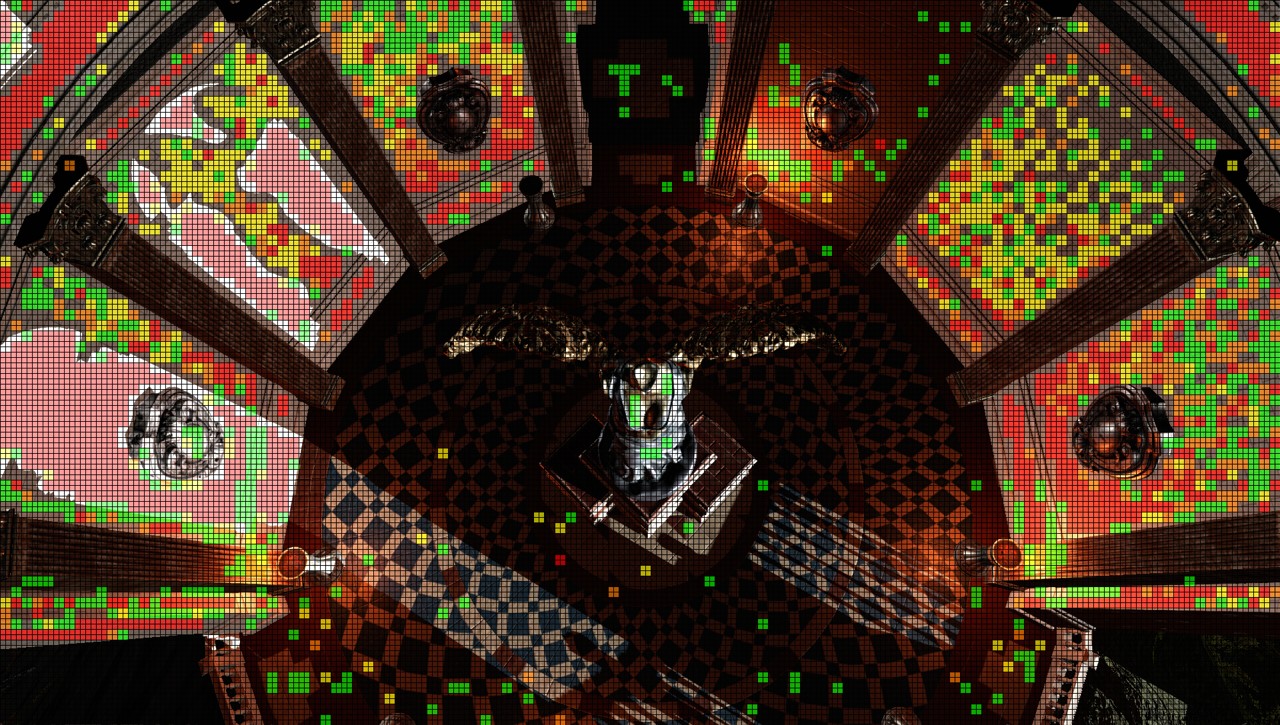}
    \caption{VRS with JNFLIP (reference)}
  \end{subfigure}
  \caption{
  Our network predicts image error metrics for real-time use cases, such as variable rate shading (VRS). The shading rate ($\square$ full, $\color{green} \blacksquare$ fine, $\color{yellow} \blacksquare$ medium, $\color{red} \blacksquare$ coarse) is selected for each image region based on a neural network's prediction from G-buffer data.
  In addition to established metrics (e.g., FLIP), we can learn Weber-corrected variants (JNFLIP) that respect perceptual context.
  }
  \label{fig:teaser}
\end{teaserfigure}
    \maketitle
    
    \section{Introduction}
The quest for more realistic and interactive rendering has led to a race for ever-increasing display resolutions and refresh rates in the hardware market. At the same time, shading costs also keep increasing due to higher software shading complexity and the intricacy of effects being used. Neither trend is expected to halt, especially considering the recent introduction of GPUs with real-time ray-tracing capabilities and the surge in popularity of virtual reality (VR) headsets, as visual quality for VR requires much higher resolutions and framerates than regular screens for the same level of perceived visual fidelity.

In light of these developments, the perceptual relevance of each pixel can change drastically depending on the view configuration and the content being viewed. To exploit this fact in real-time applications, both software and hardware solutions have recently been proposed for dynamically and locally reusing rendering information from previous frames \cite{mueller2021temporally} or changing the shading resolution across the screen depending on the displayed content \cite{yang2019adaptive, drobot2020software}.
However, the key question then becomes: how does one choose how to render each region of the screen, not knowing the end result of the shading operations? The status quo is to use recent, reprojected renderings to estimate perceptual error metrics for selecting between different (reduced) rendering modes. This comes with two limitations:
\begin{enumerate}
\item Only metrics that can be estimated from previous renderings in a computationally efficient manner may be used. This rules out most image metrics, with simple estimates becoming the norm among perceptual problems.
\item Estimation is only possible for previously seen content. The higher the amount of motion in the scene (and thus the frequency of disocclusion of previously unseen regions), the smaller the impact of these methods becomes.
\end{enumerate}
In this work, our goal is to enable the use of arbitrary metrics in real-time applications and their efficient prediction, even for previously unseen regions of the scene resulting from, e.g., fast camera movement. We present a convolutional neural network (CNN) that takes as input both reprojected renders, similarly to previous work and current-frame screen-space information that is often readily available in G-buffers before final shading, such as material properties or light visibility buffers.
We demonstrate our approach by applying our network to solve the broad problem of adaptive rendering mode selection: given a viewport that is divided into equally-sized tiles, select the suitable fidelity mode for each one.
Possible examples in current hardware include variable-rate shading (VRS), software multi-sampling, temporal shading reuse, and hybrid rendering. By enabling consistent prediction of arbitrary metrics on the entire screen regardless of scene motion, we also open the door for new methods, use cases, and perceptual metrics to appear in a real-time context.

Metric prediction for seen and unseen regions as a learning effort confronts us with novel challenges: balanced selection of training samples becomes non-trivial since conventional data preparation methods cannot be applied. Furthermore, for many practical use cases (including VRS), perceptually correct threshold values may be required, which cannot be measured for unseen regions. In this paper, we present solutions to these challenges. As a proof of concept, we use our approach to implement content-adaptive VRS. Our main contributions are as follows:
\begin{enumerate}
    \item A compact CNN for learning and predicting error metrics in real-time applications for seen and unseen regions.
    \item Two metric transforms to produce a more balanced training loss that easily generalizes for new metrics and scenes.
    \item Applying a correction to metrics that removes the need for explicitly measuring perceptual thresholds, embedding them into the trained models' predictions.
    \item Analysis and discussion of which current-frame screen-space data is most valuable for predicting error metrics.
    \item An evaluation of achievable quality, performance, and ability to generalize our learning-based approach for VRS with the current state of available hardware support.
\end{enumerate}
In the following, Section \ref{sec:related} lists previous work and necessary context to frame our contributions. Section \ref{sec:predict} describes our network and how to train it to consistently achieve high-accuracy image-error estimation in real-time. Section \ref{sec:rendering} describes how to use the network in the context of adaptive rendering mode selection, including a concrete example for application to VRS (see Figure \ref{fig:teaser}). Finally, Section \ref{sec:eval} considers the performance and quality aspects of our approach and provides an analysis of the obtained results.
    \section{Related Work} \label{sec:related}
Methods for reducing the amount of final shading computation required per display pixel are not a new concept. Mixed-resolution shading \cite{shopf2009mixed, yang2008geometry} renders expensive and low-frequency shading components at low resolution and bilaterally upsamples the results to full resolution. Decoupled shading \cite{clarberg2013sort, liktor2012decoupled, ragan2011decoupled} separates the shading rate from the visibility sampling rate by establishing a mapping between the visibility samples and shading samples and sharing or reusing shading samples where possible. Texture-space shading \cite{andersson2014adaptive, burns2010lazy, clarberg2014amfs, hillesland2016texel} computes shading in texture or patch space in an appropriate density controlled by the mip level. These software-based techniques are available for use on a wider variety of hardware but require more complicated implementation and maintenance due to their significant deviation from the hardware rasterization pipeline. 

Variable-rate shading (VRS) does not suffer from these issues. VRS can be seen as a generalization of multi-sample anti-aliasing, by which a single shading operation can be used to color not only multiple samples within a single pixel but multiple pixels. Software-based VRS implementations commonly divide the screen into $n \times n$ pixel tiles (where $n$ is an integer number) and assign shading rates---the ratio of actual pixels to the number of shading operations---independently to each tile. Current hardware implementations are even more specific and operate on $16\times16$ tiles, with a fixed set of possible shading rates \cite{turing, gen11}.
Some use cases for VRS have been targets of growing interest, such as foveated rendering \cite{vrsfoveatednvidia, franke2021time, tursun2019luminance}, a technique which uses eye-tracking hardware to direct rendering resources to the region the user focuses on \cite{patney2016towards}, or lens-optimized shading \cite{kraemer2018accelerating, yang2019presentation}, which aims at warping screen space to more closely match the final lens-corrected image \cite{unrealmatchleans}. However, these techniques are only usable with specific peripherals, such as a VR display with eye-tracking capabilities, and do not take advantage of scene-dependent information.

Content-adaptive shading, first proposed by \citeauthor{yang2019adaptive} \shortcite{yang2019adaptive}, provides a more general solution that is usable in the rendering of any 3D scene. It does so by dynamically varying the shading rate across the screen according to the perceivable level of detail of the content being rendered: the rendering result of the previous frame and the previous shading rate choices are reprojected into the current screen space and used as cues to choose the required shading rate. \citeauthor{drobot2020software} \shortcite{drobot2020software} developed a variant of this concept, designed with software-based VRS in mind.
\citeauthor{mueller2021temporally} \shortcite{mueller2021temporally} showed that shading information from previous frames can be reused for quite some time if properly sampled. \citeauthor{jindal2021perceptual} \shortcite{jindal2021perceptual} proposed a more elaborate VRS specific metric that adapts to known texture IDs. However, these techniques share several common limitations: First, they rely solely on analyzing the content from previous frames. Thus they are unable to make predictions where reprojection data isn't available. Further, they are unable to make any predictions regarding how a surface's light response or texture aliasing might change over time, which can be especially problematic with visual edges, shiny and animated materials. Finally, due to the constraints of real-time rendering, image quality needs to be measured using a computationally efficient estimator, and some form of Just-Noticeable-Difference (JND) \cite{tan2015computational} threshold. Thus, these methods have to rely on multiple approximations, leading to imprecise shading-rate decisions, which, in theory, could accumulate error over time. In practice, adaptive shading is only used after significant engine- and scene-specific tuning, such as ensuring it is only enabled in highly diffuse materials.

There has been a large amount of work in developing image metrics capable of replicating human perception, which remains inaccessible in real-time environments. \citeauthor{flip} \shortcite{flip} presented the FLIP estimator, inspired by models of the human visual system and designed with particular focus on the differences between rendered images and corresponding ground truths.  \citeauthor{zhang2018unreasonable} \shortcite{zhang2018unreasonable} discovered that, during image classification, the intermediate image representation produced by the network could be used for comparison with other images. \citeauthor{wolski2018dataset} \shortcite{wolski2018dataset} created a data set of image pairs with user markings of where they perceive distortions and a convolutional network trained on it capable of predicting markings in new images. 
There has also been a surge in the development of deep learning approaches for the post-processing of real-time renderings, such as super-resolution and temporal anti-aliasing of rasterized surfaces \cite{dlss, thomas2020reduced}, or denoising of ray-traced ones \cite{meng2020real, hasselgren2020neural}.
    \begin{figure}
  \centering
  \includegraphics[width=0.85\linewidth]{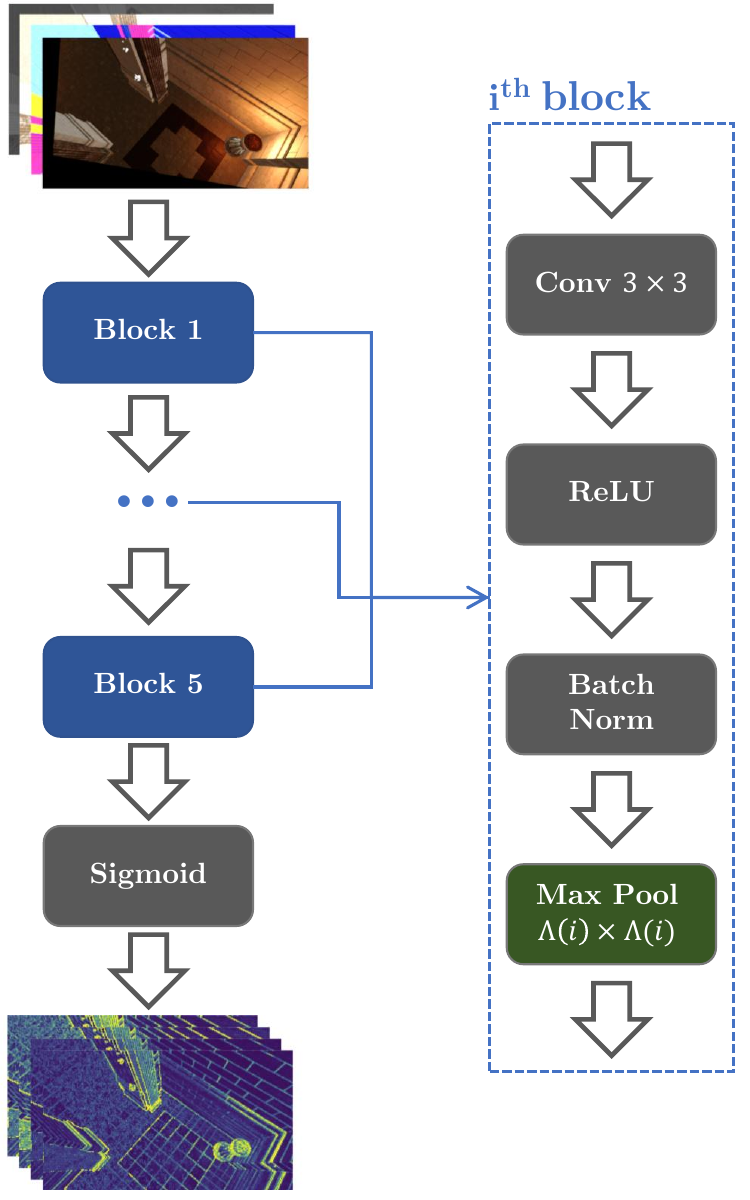}
  \caption{Proposed network architecture for metric prediction. Each block $i$ performs down-pooling at size $\Lambda(i)$.}
  \label{fig:network}
\end{figure}

\section{Metric Prediction} \label{sec:predict}
Conventionally, a reference image metric $f(I, J)$ computes the perceptual difference between a reference image $I$ and a candidate image $J$. No-reference methods $f(J)$ guess perceptual issues given the expected proprieties and common distortions in natural images. Values may be computed for the entire image domain or regions thereof.
In this work, we aim instead to estimate $f(I, I')$, where $I'$ represents an informed approximation of the reference $I$, such as a lower-resolution rendering of $I$.
Our goal is to predict $f(I, I')$ directly, without explicitly computing either $I$ or $I'$ by exploiting other, more easily available screen-space scene information instead.

Our deep learning-based approach enables fast prediction of complex metrics that would otherwise incur significant computational overhead. However, one challenge to overcome is the sensitivity of machine learning to unbalanced training data sets; another is that the practical applications of $f(I, I')$ often involves spatially varying parameters, e.g., the local just noticeable difference (JND) at each point in $I$~\cite{yang2019adaptive}.
In this section, we introduce our network architecture, discuss which input data should be used to predict metrics, and present our solution to the output imbalance problem. Furthermore, we show how the spatially varying JND threshold can be integrated directly into the trained model. 
For the sake of brevity, the visual illustrations of our approach will focus exclusively on the example of predicting the error when $I$ and $I'$ vary in shading rate. In the figures displayed in this article, plotted or color-coded values of $f(I,I')$ show the difference between reference $I$ and corresponding $I'$ obtained with coarser $2\times2$ shading rate for a given metric $f$.

\subsection{Convolutional Network Architecture} \label{sec:network}
Figure \ref{fig:network} shows the schematic of our convolutional network architecture. It consists of $3\mult3$ convolutions, interlaced with rectified linear units (ReLU) and batch normalization. 
We optimize for prediction performance by pooling as early as possible in the network and maintain a consistent amount of parallelism by dividing hidden channels into independent groups at the same rate at which down-pooling is performed---that is, we try to keep the number of independent groups times the number of pixels remains the same. A single final sigmoid layer is used to constrain the output to the range $[0, 1]$.
To support optimized generation of (conservative) predictions for arbitrarily-sized image regions (e.g., for application to hardware VRS), maximum pooling is done depending on an intended region size $w \mult w$ (for per-pixel predictions, $w = 1$). 
The size $\Lambda(i)$ in down-pooling layer $i$ is:
\begin{align}
   \Lambda(i) = \begin{cases}
        2 &\text{if } \frac{w}{2^i} > 2 \> \text{ and } i < 5\\
        \lceil\frac{w}{2^i}\rceil &\text{otherwise}
    \end{cases}
\end{align}
The design of our network is governed by its intended use in real-time applications: given sufficient training time, the network is capable of learning sophisticated features while prediction remains fast. Its layout makes it compatible with optimized, massively parallel inferencing solutions, such as TensorRT.
Furthermore, the per-region predictions for $w > 1$ can be passed on directly to tile-based procedures.

We converged on our eventual design after comparing more complex alternatives, which all underperformed or provided no visible benefit over the simpler solution. These alternatives included using partial convolutions---with and without data masking---and rendering-aware denormalization. We also decided on maximum pooling as it provided higher accuracy than downscaling purely through convolution.

\begin{figure}
  \centering
  \begin{minipage}{0.945\columnwidth}
  \begin{subfigure}{0.49\textwidth}
    \includegraphics[width=\textwidth]{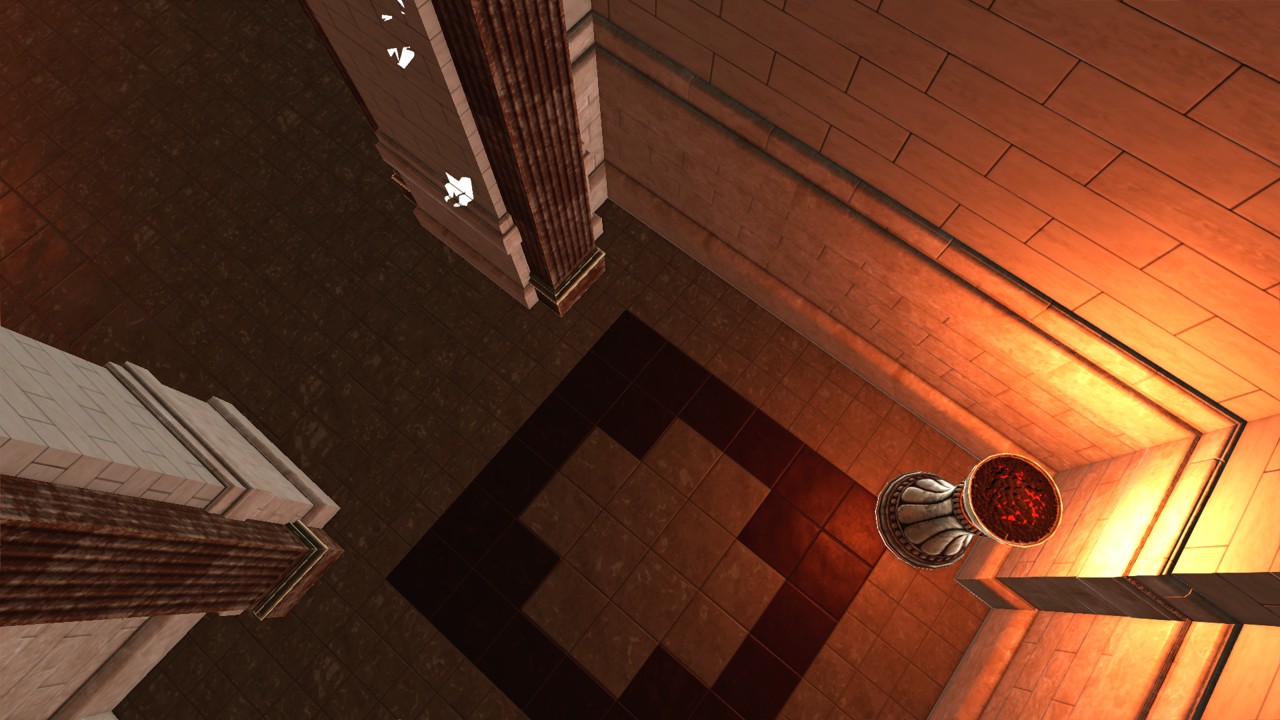}
    \caption{Reference}
  \end{subfigure}
  \hfill
  \begin{subfigure}{0.49\textwidth}
      \includegraphics[width=\textwidth]{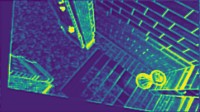}
      \caption{Temporal Reprojection}
  \end{subfigure}
  \hfill
  \begin{subfigure}{0.49\textwidth}
      \includegraphics[width=\textwidth]{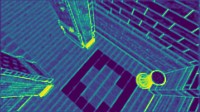}
      \caption{Surface Data}
  \end{subfigure}
  \begin{subfigure}{0.49\textwidth}
      \includegraphics[width=\textwidth]{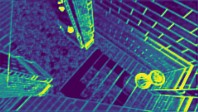}
      \caption{Both}
  \end{subfigure}
  \hfill
  \begin{subfigure}{0.49\textwidth}
      \includegraphics[width=\textwidth]{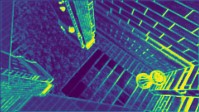}
      \caption{Optimal 4 Channels}
  \end{subfigure}
  \hfill
  \begin{subfigure}{0.49\textwidth}
      \includegraphics[width=\textwidth]{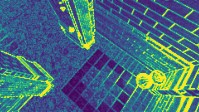}
      \caption{Target}
  \end{subfigure}
  \end{minipage}
  \hfill
  \begin{minipage}{0.0436\columnwidth}
    \raisebox{1.6em}{\includegraphics[width=\textwidth]{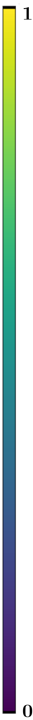}}
  \end{minipage}
  \caption{Network predictions for FLIP error between the full-resolution reference image and a coarser, $2\times2$-shaded versions. Results were obtained from networks trained with different screen-space input sets. Number of input channels used are 3, 16, 19 and 4 in (b), (c), (d) and (e), respectively.}
  \label{fig:channels}
\end{figure}

\subsection{Input Data}
\label{sec:input_data}
Our solution aims to leverage as input any screen-space information that becomes available in real-time rendering pipelines prior to expensive stages that can benefit from accurate metric predictions. Hence, it presents an ideal fit for ubiquitous deferred shading pipelines, which provide a range of screen-space information via the G-buffer. Outputs of previous frames are also commonly obtained as a byproduct of rendering or at little additional cost through temporal reprojection.
The question then becomes which of these resources to choose as inputs for the network to yield high accuracy while keeping the input set compact. 
We assessed commonly available G-buffer contents and statistically analyzed how influential each is on the prediction of perceptual error metrics.
Our reference rendering pipeline uses deferred shading, with cascading shadow maps, screen-space ambient occlusion, fast approximate anti-aliasing, and tone mapping with automatic exposure selection. The pipeline was implemented on top of Falcor \cite{falcor} and the network trained on established ORCA scene assets (Amazon Bistro \cite{ORCAAmazonBistro} and Unreal Engine 4's Suntemple \cite{OrcaUE4SunTemple}).

We found that directly available information in the G-buffer---such as view-space normals, diffuse color, or roughness---enables reasonable predictions across the entire screen. However, it lacks a myriad of information that otherwise would have to be explicitly encoded, such as lighting, tone mapping, or other effects. As shown in Figure \ref{fig:channels}, we found the temporal reprojection of final color from previous frames to be a valuable asset (similar to \cite{yang2019adaptive} and \cite{drobot2020software}), as it contains most of this missing information. However, color reprojection is spatially limited to previously seen regions only and thus presents decreasing benefits in use cases with more obstructions, animated scenes or fast-paced camera movement. Figure \ref{fig:channels} proves that using temporal reprojection with a quickly changing view or scene does not suffice to produce adequate predictions for the current frame. Hence, a good prediction solution should weight available inputs differently, depending on whether it is predicting for recently seen or newly disoccluded, unseen regions. We assumed (and experimentally confirmed) that the network's prediction quality is highest if reprojected color is paired with a binary mask (seen = 1, unseen = 0).

\begin{table}
\centering
  \caption{DeepLIFT contribution of network inputs (FLIP)}
  \label{tab:input}
  \begin{tabular}{lccc}
    \toprule
    Channels & Format & Seen Regions & Unseen\\
    \midrule
    Reprojected Color & RGB & 31.37\% & ---\\
    Reprojection Mask & Bool & 17.26\% & 8.67\%\\
    View Normals & RGB & 16.89\% & 33.63\%\\
    Diffuse Color & RGB & 14.8\% & 42.59\%\\
    View Normal Z & Float & 10.12\% & 20.15\%\\
    Shadowing & Float & 7.48\% & 9.36\%\\
    Roughness & Float & 5.41\% & 6.77\%\\
    Specular Color & RGB & 5.73\% & 10.61\%\\
    Reflect Product & Float & 1.06\% & 1.33\%\\
    Emissive Color & RGB & 0.01\% & 0.03\%\\
  \bottomrule
\end{tabular}
\end{table}

To quantify the contribution of each input candidate, we used DeepLIFT \cite{shrikumar2017learning,ancona2017towards} on a model trained on all pre-selected candidate inputs and computed attribution scores on a large validation set from our test suite. Table \ref{tab:input} lists the mean absolute attribution score of each candidate input, as identified on the FLIP metric. As expected, reprojected color contributes the most, but even more so if masked (accepted if previously seen, zero otherwise). The contribution of diffuse material colors is highest for unseen regions. Other inputs are less important, such as emissive material color, for which we found no anecdotal or statistical benefit, or the dot product between the surface normal and the reflection vector, which is redundant if view-space normals are provided directly. Most RGB channels are relatively redundant, with whichever channel being first in the input order becoming the dominant one and representing the majority of the accuracy of the whole group. The only exception was normals, where the Z-axis is always the dominant one. We also found no advantages of training with HDR for RGB input data instead of 8-bit color channels. Using this knowledge, we can derive effective yet compact input data sets. For real-time applications, we propose to use a single 4-channel texture containing the reprojection mask, one RGB channel (any) of the reprojected color, one RGB channel of diffuse material color, and the Z-axis of the view-space normals. This provides a good tradeoff between desired low inference time and prediction quality since these four account for $52.08\%$ of the network's prediction capability, according to DeepLIFT.

\subsection{Reparameterization} 
For a given perceptual metric, its output value distribution can change drastically with different environments and rendering settings.
We noticed in our experiments that, for most metrics, the tested scenes produced mostly low output values and only a few very high outliers. Such an unbalanced target distribution might prevent the network from converging to a reasonable solution altogether when trained on arbitrary scenes. In theory, this problem becomes less noticeable the more data and a greater variety of scenes are provided. However, our goal is to provide a solution that can be efficiently trained with a limited training set, as well as arbitrary metrics, scenes, and rendering settings, yet still, generalize across them well.

We note that for many real-time applications, high metric prediction accuracy is most relevant within a limited range of values that drive performance-related optimizations, such as render mode selection.
Thus, we choose to tackle the data imbalance issue by using a modified parameter space that balances the data distribution while preserving the relevant information in it.

Let $\mathcal{L}(Y, \hat{Y})$ be a given loss function, where $\hat{Y} \in [0, 1]$ is a set of predicted values in transformed space, and $Y = f(I, I') \in [0, 1]$ the corresponding target values. We define a new loss function that measures the difference between predictions $\hat{Y}$ and targets $Y$ after transforming them to a new parameter space according to a function $\mathcal{T}$:
\begin{equation}
    \mathcal{L}_{adaptive}(Y,\hat{Y}) = \mathcal{L}(\mathcal{T}(Y), \hat{Y})
\end{equation}
We then use mean absolute error (MAE) as our $\mathcal{L}$ loss function:
\begin{equation}
    \mathcal{L}_{MAE\;adaptive}(Y,\hat{Y}) = \frac{1}{n} \sum_i^N{|\mathcal{T}(Y_i) - \hat{Y}_i|}
\end{equation}
If predictions in non-transformed space are required (e.g., for comparison with perceptual thresholds), they can be obtained as $\mathcal{T}^{-1}(\hat{Y})$. In the following, we describe our two proposed different reparameterization transforms.

\begin{figure*}
  \centering
  \begin{subfigure}{0.196\textwidth}
    \includegraphics[width=\textwidth]{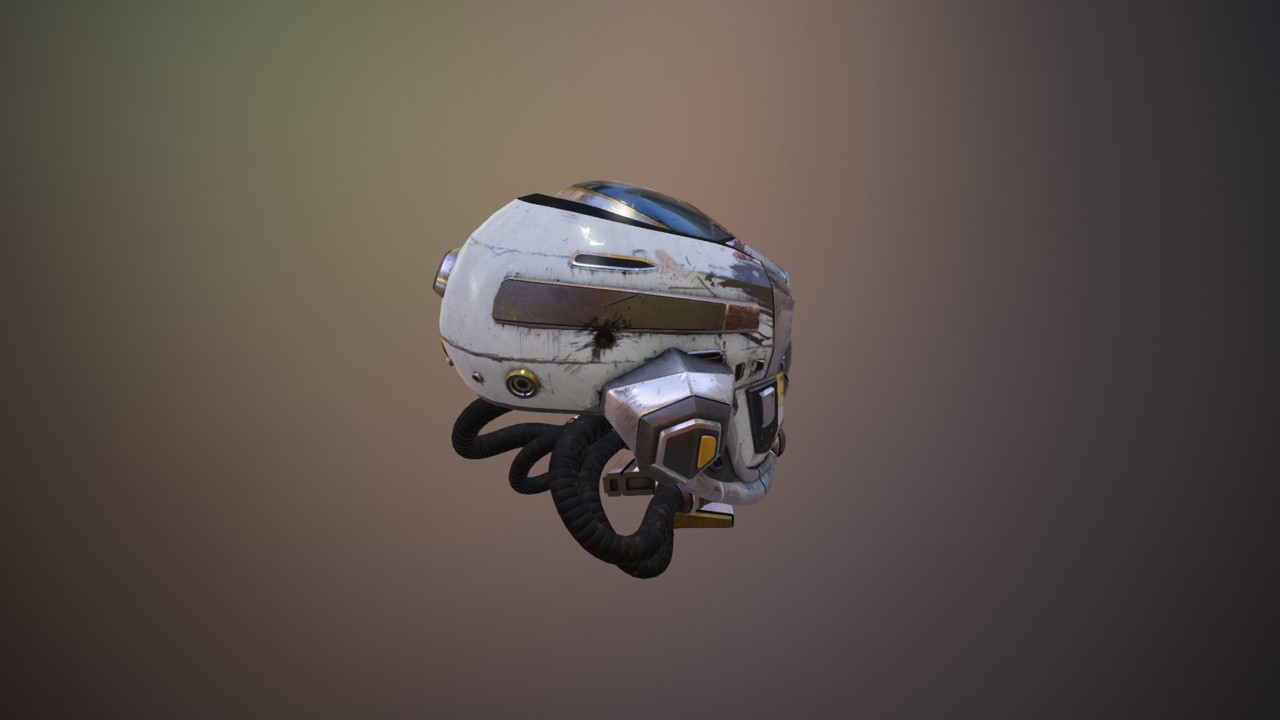}
    \caption{Helmet scene}
  \end{subfigure}
  \hfill
  \begin{subfigure}{0.196\textwidth}
    \includegraphics[width=\textwidth]{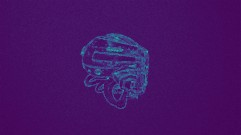}
    \caption{Ground truth}
  \end{subfigure}
  \hfill
  \begin{subfigure}{0.196\textwidth}
    \includegraphics[width=\textwidth]{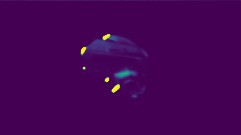}
    \caption{MSE Loss}
  \end{subfigure}
  \hfill
  \begin{subfigure}{0.196\textwidth}
    \includegraphics[width=\textwidth]{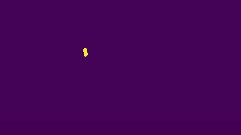}
    \caption{MAE Loss}
  \end{subfigure}
  \hfill
  \begin{subfigure}{0.196\textwidth}
    \includegraphics[width=\textwidth]{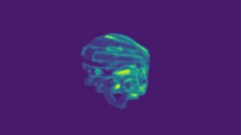}
    \caption{Clamped MAE (ours)}
  \end{subfigure}
  \caption{Example of the network predicting FLIP on an extremely unbalanced scenario \cite{helmet}, containing mostly background and highly reflective surfaces. Training this network to predict this error (b) with traditional losses causes it to underestimate the metric (c,d). Applying our transform on the parameter space remedies the issue (e).}
  \label{fig:helmet}
\end{figure*}

\subsubsection{Clamped Transform}
A computationally efficient but lossy reparameterization solution is to re-scale the metric, so its output distribution is centered at $0.5$, and clamp outlier values to $[0,1]$ ---existing work on HDR imagery shows us this is not unreasonable \cite{lee2018deep}.
Let $Y_i$ be a value to be transformed and $\mu_Y$ the mean value of all target values in the data set. We define the clamped transform:
\begin{align}
    \mathcal{T}_{clamped}(Y_i) = \max(\min(\frac{Y_i}{2 \mu_Y}, \> 1), \> 0)
\end{align}
where $\mu_Y$ can either be precomputed before training or estimated on-the-fly as a running average of previously seen values for $Y$. We found this transform to improve prediction efficacy on all of our tests, exemplified by Figure \ref{fig:helmet}, and suggest it as the default choice. Note that due to its lossy nature, $\mathcal{T}^{-1}$ only exists in the $[0, 1]$ range.

\subsubsection{Logistic Transform}
Due to its assumptions, the clamped transform may fail to generalize in special cases. This could occur when training on a data set with higher values---and thus, higher $\mu_Y$---or if using a metric with a vastly different distribution. Additionally, it also removes information and zeroes out derivatives for the high outlier values. 
In cases where this becomes an issue, we propose using instead a transform based on the logistic function $\mathcal{S}$, which is a bounded function with a bell-shaped derivative that is defined everywhere and has its peak at the curve's midpoint. We use the standard logistic function, centered on $\mu_Y$:
\begin{equation}
    \mathcal{S}(Y_i) = \frac{1}{1 + e^{-k( Y_i-\mu_Y )}}
\end{equation}
This function allows us to re-center the prediction distribution for any value of $\mu_Y$, while establishing increased importance of accuracy for values near $\mu_Y$, without zeroing any derivatives. Furthermore, it allows adjusting the relative impact of outliers using the logistic growth rate hyperparameter $k$. In practice, we found that $k=10$ works best across the evaluated data sets. $\mathcal{S}$, as defined above, does not produce values $\in [0,1]$.
Hence, we normalize it as:
\begin{equation}
   \mathcal{T}_{logit}(Y_i) = \frac{\mathcal{S}(Y_i) - \mathcal{S}(0)}{\mathcal{S}(1)  - \mathcal{S}(0)}
\end{equation}
\subsection{Weber Correction}\label{sec:weber}
Several use cases of real-time perceptual metrics, such as content-adaptive shading \cite{yang2019adaptive, drobot2020software} or decoupled shading \cite{mueller2021temporally} use the just-noticeable difference (JND) threshold to inform performance decisions, like render mode selection.
However, state-of-the-art approaches rely on explicit computation or reprojection to obtain this---spatially varying---value.
Hence, it is only available \emph{after} rendering or for previously seen regions, but not for unseen regions before shading. 
We solve this issue for our learning-based approach by embedding the visual component of the prediction directly into the model.
To enable visually-based decisions, for the current frame, we must estimate the final image error $E$ and compare it with the JND threshold $\mathcal{W}$. 
Based on Weber’s law \cite{tan2015computational}, \citeauthor{yang2019adaptive} \shortcite{yang2019adaptive} define this threshold $\mathcal{W}_i$ and its applied relation to the visual image error $E_i$ at each location $i$ as:
\begin{equation}
    E_i \le \mathcal{W}_i = t \cdot (L_i + l)
\end{equation}
where $L_i$ is the average luminance at location $i$, $t$ is a user-selected sensitivity threshold, and $l$ is the environment luminance, which affects the sensitivity to dark areas. This definition is only valid assuming a metric whose output values $E$ directly represent visual error on a luminance scale (e.g., FLIP). The relation is equivalent to:
\begin{equation}
    \frac{E_i}{L_i + l} \le t
\end{equation}
Hence, instead of computing the perceptually-corrected threshold in real-time, we can train the network to estimate an already corrected metric $Y'$, enabling the model to specialize for its eventual real-time application and reducing computational cost at runtime:
\begin{equation}
    Y' = \frac{E}{L + l}
\end{equation}
Our experiments include Weber-corrected variants derived from existing metrics: just-noticeable FLIP (JNFLIP) and the just-noticeable variant of the image error estimation used in \cite{yang2019adaptive} (JNYang).

    \section{Real-Time Render Mode Selection} \label{sec:rendering}
\begin{figure*}
  \centering
  \includegraphics[width=\linewidth]{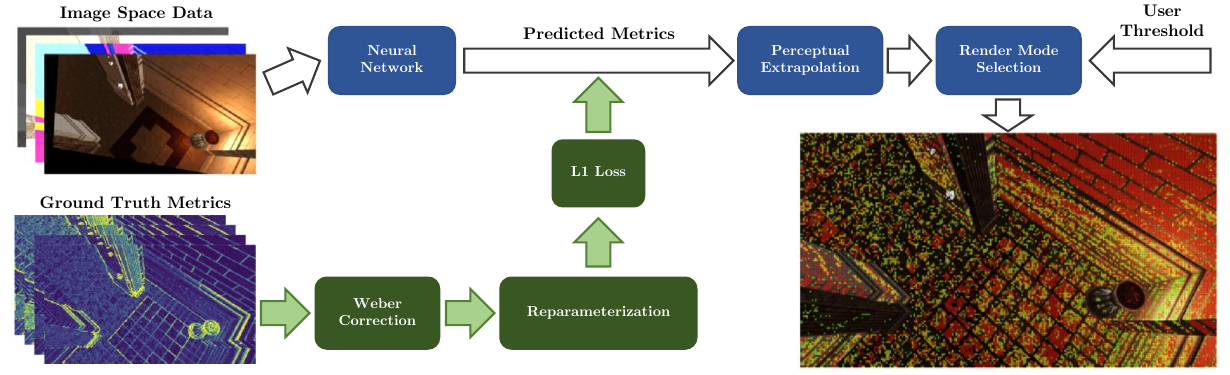}
  \caption{Proposed rendering mode selection pipeline for content-adaptive shading with VRS. Inputs are provided to a network that has been trained to predict a perceptual metric, with Weber correction and reparameterization applied. At run-time, the network predicts the implied image error for selecting different shading rates. Based on these predictions and a user-defined threshold, the final shading rate ($\blacksquare$ full, $\color{green} \blacksquare$ fine, $\color{yellow} \blacksquare$ medium, $\color{red} \blacksquare$ coarse) is selected for each image region.}
  \label{fig:rendering}
\end{figure*}

We describe how our metric prediction network can be used for render mode selection.
Furthermore, we describe optimizations for applying it to content-adaptive shading with VRS (see Figure \ref{fig:rendering}).

\subsection{Data Capture}
In order to train a metric prediction network for render mode selection, capturing of training data should be performed with the same render engine that the model is intended to be used with eventually. In the case of a perceptual metric, reference images for different rendering modes should be captured only after all image post-processing and effects have been applied since the computed errors should capture the perceived visual difference.

For data collection and training, we start by capturing the environments at representative viewpoints at each possible render mode. This is necessary for generating the training and validation targets of any metric that relies on $I,I'$ image pairs. We then compute the metric between each render mode and the reference image obtained without any optimizations active.
We also capture the corresponding network input data for each rendered frame, both temporarily reprojected and from the current frame.

\subsection{Mode Inference}
Rather than predicting render modes directly, we suggest producing a continuous error prediction and perform mode selection based on user-defined thresholds, e.g., the JND threshold, as this allows for greater control by artists and application users alike.
Consequently, we can exploit our metric prediction network for this task. We set the layout for our network such that the predicted metric between a render mode and the reference image is computed in a separate output channel for each available mode.  We can therefore iterate these channels in order of increasing computational cost and check if any presents a perceptual loss lower than the defined threshold. If no available channel presents an acceptable value for a given tile, we apply the highest quality mode instead:

\begin{lstlisting}[language=Python]
chooseMode(metric, tile)
  for each mode in increasing cost
    if metric[mode, tile] < threshold
        return mode
  return reference mode
\end{lstlisting}

\subsection{Rate Extrapolation for VRS} \label{sec:extrapolation}
Many modern real-time graphics solutions offer support for VRS, which allows selecting different shading rates for individual objects or image regions to economize on expensive fragment shader invocations.
Commonly supported shading rates include fundamental squares ($1\times1$, $2\times2$ and $4\times4$) and rectangles with conforming side lengths.
For this particular use case, the metric values for similar shading rates are strongly correlated: 
similar to \citeauthor{yang2019adaptive} \shortcite{yang2019adaptive}, we can reduce the number of output channels by extrapolating the outputs of multiple channels from just a few.
Let $\hat{Y}_{u \mult v}$ be an output channel of the network, where $u$ and $v$ are its corresponding horizontal and vertical shading strides, respectively. Let $k = 2.13$ capture the constant relative change in error when switching from a shading rate to its half (e.g., $2\times2 \rightarrow 4\times4$), as derived by \citeauthor{yang2019adaptive} \shortcite{yang2019adaptive}. We can approximate lower shading rates from higher ones, allowing for using only two output channels---the network predictions for $1\times2$ and $2\times1$ shading rates:

\begin{equation} \label{eq:extrapolation}
   \hat{Y}_{u \times v} \approx \begin{cases}
        \max(\hat{Y}_{\frac{u}{2} \times v}, \hat{Y}_{u \times \frac{v}{2}}) &\text{if } u = v\\
        \max(\hat{Y}_{\frac{u}{2} \times \frac{v}{2}} \cdot k, \hat{Y}_{\frac{u}{2} \times v}) &\text{if } u > v\\
        \max(\hat{Y}_{\frac{u}{2} \times \frac{v}{2}} \cdot k,  \hat{Y}_{u \times \frac{v}{2}}) &\text{if } u < v
    \end{cases}
\end{equation}

The values for shading rate $2\times2$ can be extrapolated from $1\times2$ and $2\times1$.
Following Equation \ref{eq:extrapolation}, $2\times4$ can further be obtained from $1\times2$ and $2\times2$, $4\times2$ from $2\times1$ and $2\times2$, $4\times4$ from $2\times4$ and $4\times2$, and so on.
We found that square rates are approximated with higher precision than non-square rates. Thus, in practice, we recommend using 4 output channels ($1\times2$, $2\times1$, $2\times4$, $4\times2$) and extrapolating the others for good quality/performance trade-off.

    \section{Evaluation} \label{sec:eval}
We evaluate our approach regarding prediction quality, performance, and robustness. We use the 4-channel input set we recommended in Section \ref{sec:input_data}, and randomly captured $12820$ viewpoint pairs to simulate camera movement in a total of 8 different scenes. Results were generated on a Windows 10 PC with an i7 CPU @ 3.40GHz, 16GB RAM, and an NVIDIA RTX 2080TI GPU.

\begin{table*}
  \caption{Prediction quality on test sets across six different scenes. The network has only been trained on the three scenes in the left column (Suntemple and Amazon Bistro). For each scene, we give the number of triangles ($\Delta$), unique materials ($\circledast$), and the achieved $R^2$ score (coefficient of determination), mean average error (MAE, i.e., the discrepancy between measured and predicted perceptual metric, both total and underestimation only) and variance ($\sigma_{MAE}$) of the total MAE.} \label{tab:validation}
  
  \begin{tabular}{c|cccc|cccc|cccc}
    \multirow{2}{*}{} & \multicolumn{4}{c|}{Suntemple, 606k $\Delta$, 48 $\circledast$}                                   & \multicolumn{4}{c|}{Amazon Bistro (Exterior), 2.8M $\Delta$, 132 $\circledast$}                    & \multicolumn{4}{c}{Amazon Bistro (Interior), 1M $\Delta$, 71 $\circledast$}                     \\ \cline{2-13} 
                      & $R^2$         & MAE$_{total}$ & MAE$_{under.}$ & $\sigma_{MAE}$ & $R^2$         & MAE$_{total}$ & MAE$_{under.}$ & $\sigma_{MAE}$ & $R^2$         & MAE$_{total}$ & MAE$_{under.}$ & $\sigma_{MAE}$ \\ \hline
    FLIP              & \textbf{90\%} & 5.99e-2       & 4.40e-2        & 7.30e-2         & \textbf{81\%} & 8.55e-2       & 4.40e-2        & 1.08e-1         & \textbf{78\%} & 7.88e-2       & 4.46e-2        & 9.87e-2         \\
    PSNR              & \textbf{92\%} & 3.15e-2       & 1.42e-2        & 3.21e-2         & \textbf{82\%} & 4.06e-2       & 1.98e-2        & 4.55e-2         & \textbf{80\%} & 3.85e-2       & 1.25e-2        & 4.71e-2         \\
    LPIPS             & \textbf{79\%} & 4.32e-2       & 2.46e-2        & 4.54e-2         & \textbf{77\%} & 4.15e-2       & 2.23e-2        & 4.51e-2         & \textbf{72\%} & 3.39e-2       & 1.52e-2        & 3.83e-2         \\
    JNYang            & \textbf{87\%} & 7.75e-2       & 5.02e-2        & 1.11e-1         & \textbf{84\%} & 7.93e-2       & 4.15e-2        & 1.29e-1         & \textbf{78\%} & 8.59e-2       & 4.72e-2        & 1.32e-1         \\
    JNFLIP            & \textbf{88\%} & 7.37e-2       & 4.78e-2        & 9.56e-2         & \textbf{82\%} & 9.52e-2       & 4.10e-2        & 8.21e-2         & \textbf{79\%} & 9.51e-2       & 4.12e-2        & 8.21e-2         \\ \hline
    \multirow{2}{*}{} & \multicolumn{4}{c|}{Emerald Square (Day), 10M $\Delta$, 220 $\circledast$}                        & \multicolumn{4}{c|}{Emerald Square (Dusk), 10M $\Delta$,  222 $\circledast$}                      & \multicolumn{4}{c}{Sibenik Cathedral, 75k $\Delta$, 15 $\circledast$}                               \\ \cline{2-13} 
                      & $R^2$         & MAE$_{total}$ & MAE$_{under.}$ & $\sigma_{MAE}$ & $R^2$         & MAE$_{total}$ & MAE$_{under.}$ & $\sigma_{MAE}$ & $R^2$         & MAE$_{total}$ & MAE$_{under.}$ & $\sigma_{MAE}$ \\ \hline
    FLIP              & \textbf{94\%} & 4.98e-2       & 2.51e-2        & 7.56e-2         & \textbf{94\%} & 6.18e-2       & 1.24e-2        & 6.99e-2         & \textbf{91\%} & 4.07e-2       & 2.50e-2        & 6.18e-2         \\
    PSNR              & \textbf{83\%} & 5.45e-2       & 3.22e-2        & 5.72e-2         & \textbf{92\%} & 4.95e-2       & 6.4e-3         & 4.30e-2         & \textbf{88\%} & 2.95e-2       & 1.15e-2        & 4.05e-2         \\
    LPIPS             & \textbf{80\%} & 4.15e-2       & 2.29e-2        & 4.55e-2         & \textbf{81\%} & 3.56e-2       & 1.38e-2        & 3.85e-2         & \textbf{70\%} & 3.28e-2       & 2.09e-2        & 3.68e-2         \\
    JNYang            & \textbf{94\%} & 5.03e-2       & 1.83e-2        & 9.50e-2         & \textbf{92\%} & 5.15e-2       & 1.55e-2        & 1.04e-1         & \textbf{86\%} & 6.61e-2       & 2.34e-2        & 9.61e-2         \\
    JNFLIP            & \textbf{90\%} & 7.30e-2       & 2.37e-2        & 8.15e-2         & \textbf{82\%} & 9.62e-2       & 0.61e-2        & 1.53e-1         & \textbf{81\%} & 9.18e-2       & 1.06e-2        & 9.17e-2        
  \end{tabular}
\end{table*}

\begin{figure*}
\setlength{\tabcolsep}{4pt}
  \centering
  \begin{tabular}{lccc}
     & Ground Truth & Seen Regions Mask & Prediction\\
    \rotatebox{90}{PSNR}
        & \includegraphics[valign=c, width=0.31\textwidth]{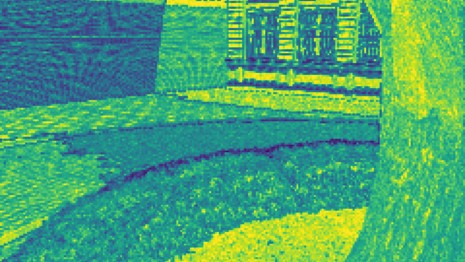}
        & \includegraphics[valign=c, width=0.31\textwidth]{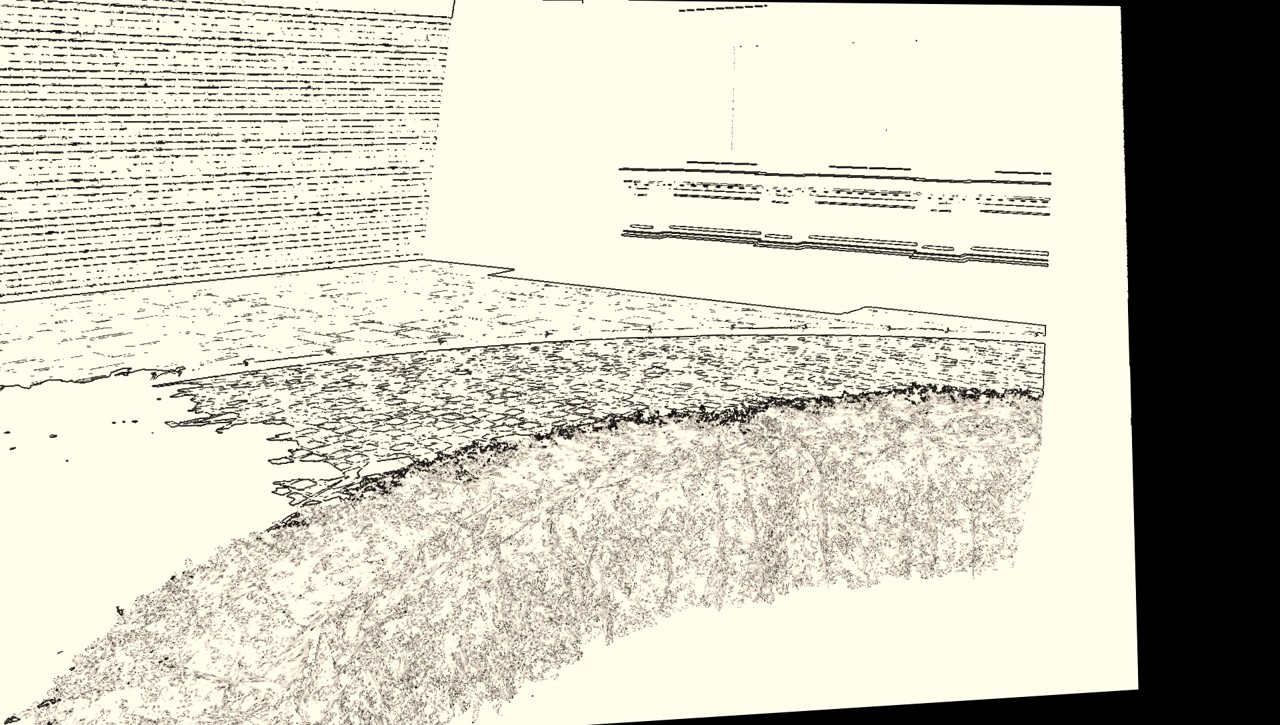}
        & \includegraphics[valign=c, width=0.31\textwidth]{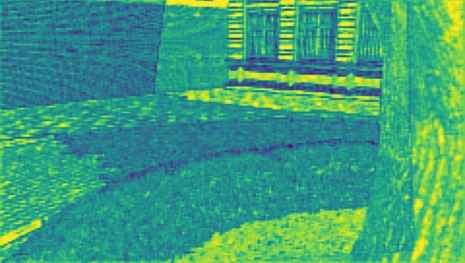} \\
    \rotatebox{90}{FLIP}
        & \includegraphics[valign=c, width=0.31\textwidth]{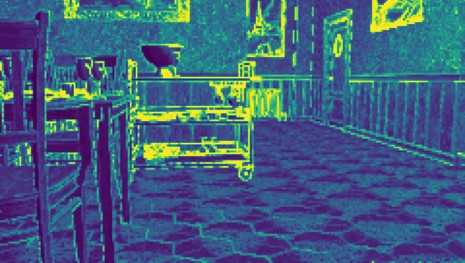}
        & \includegraphics[valign=c, width=0.31\textwidth]{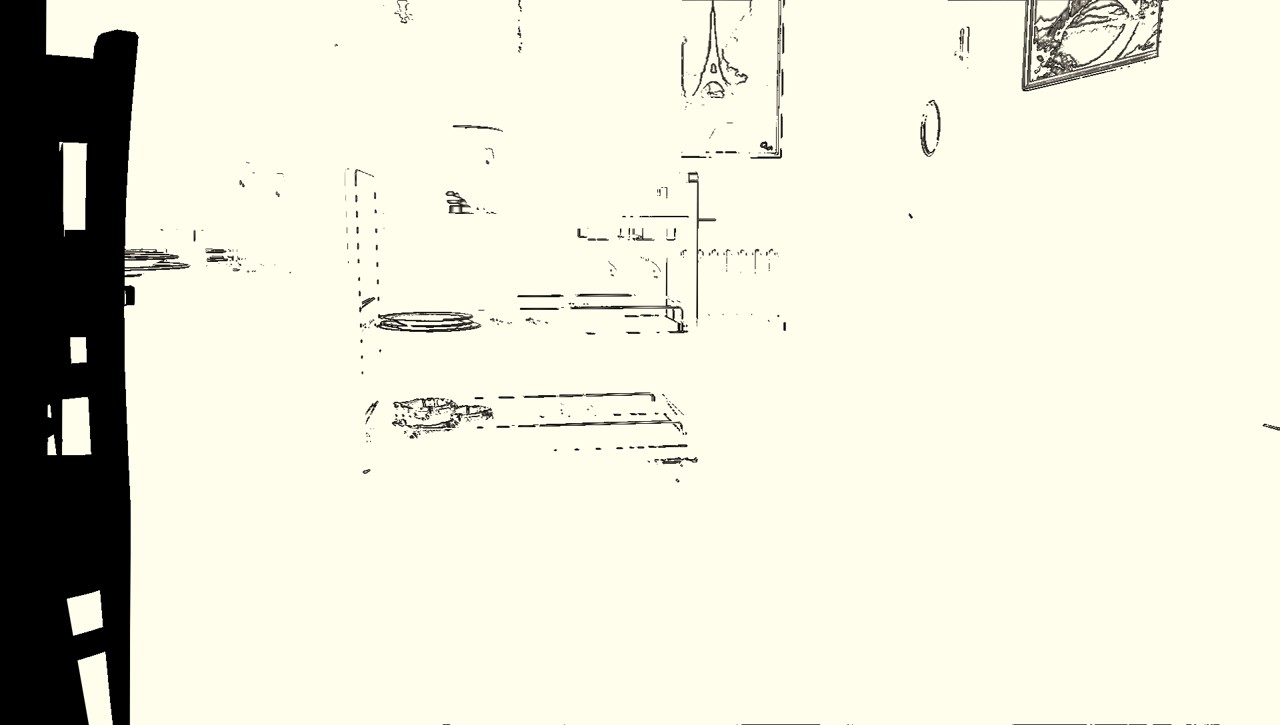}
        & \includegraphics[valign=c, width=0.31\textwidth]{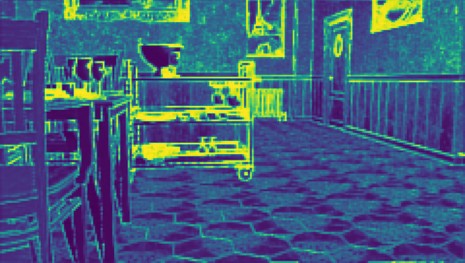} \\
    \rotatebox{90}{LPIPS}
        & \includegraphics[valign=c, width=0.31\textwidth]{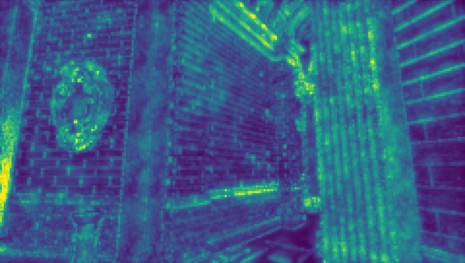}
        & \includegraphics[valign=c, width=0.31\textwidth]{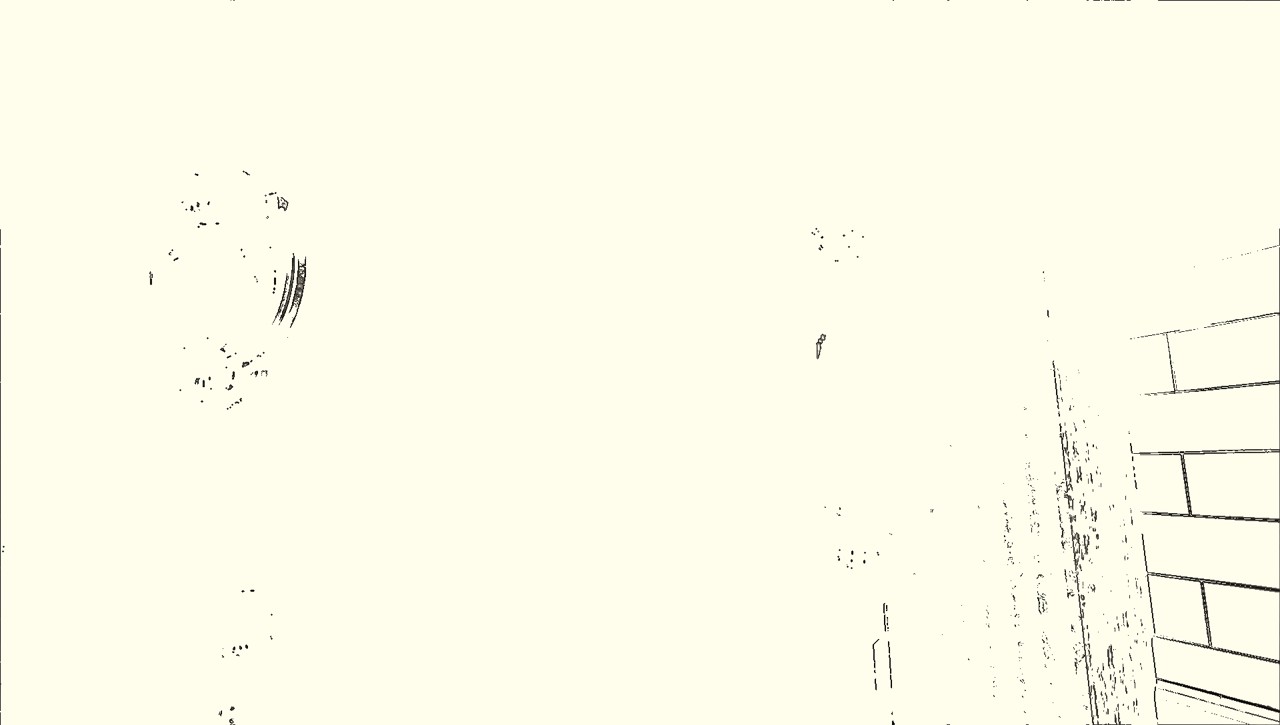}
        & \includegraphics[valign=c, width=0.31\textwidth]{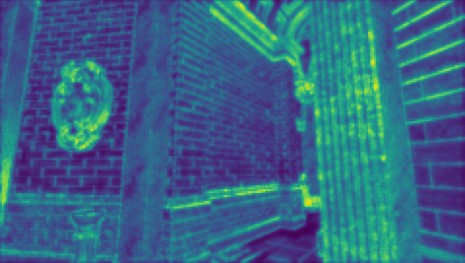} \\
    \rotatebox{90}{JNFLIP}
        & \includegraphics[valign=c,  width=0.31\textwidth]{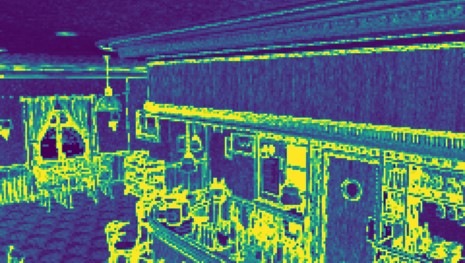}
        & \includegraphics[valign=c, width=0.31\textwidth]{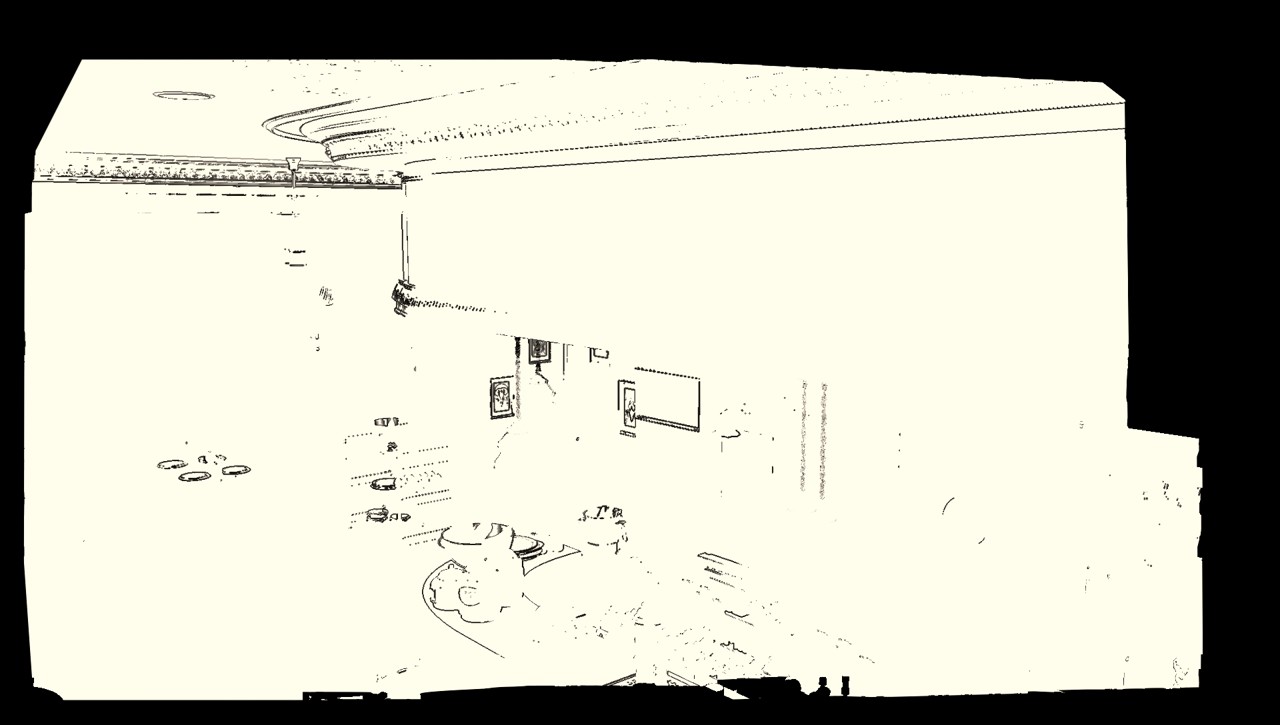}
        & \includegraphics[valign=c, width=0.31\textwidth]{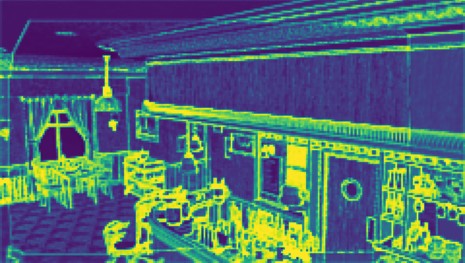}\\
    \rotatebox{90}{JNYang}
        & \includegraphics[valign=c, width=0.31\textwidth]{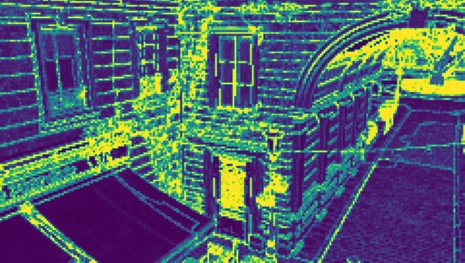}
        & \includegraphics[valign=c, width=0.31\textwidth]{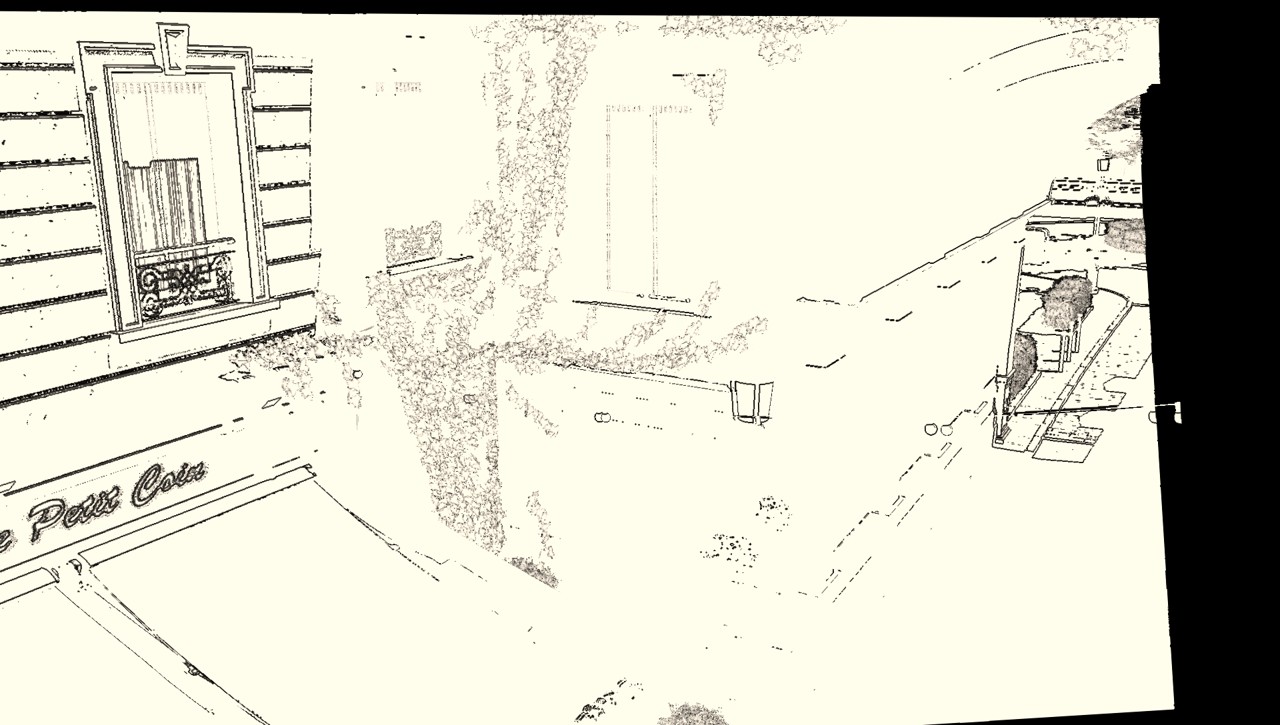}
        & \includegraphics[valign=c, width=0.31\textwidth]{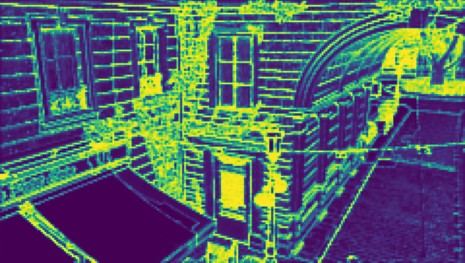} \\ 
  \end{tabular}
  \caption{Examples of our network predicting metrics in tested scenes. Black in the center column indicates unseen regions in the current frame. All metrics performed similarly across tested scenes, with no obvious outliers or catastrophic failures.}
  \label{fig:metric_scenes}
\end{figure*}

\subsection{Metric Prediction}
To evaluate the network's prediction capability, we trained and tested it with three established error metrics (PSNR, FLIP, and LPIPS), as well as the Weber-corrected variants (JNFLIP and JNYang). Validation was performed for each scene from Section \ref{sec:input_data}, using $64$ random viewpoint pairs that were withheld during training, as well as on three scenes the network was never trained on: Emerald Square (day/dusk) \cite{ORCANVIDIAEmeraldSquare} and Sibenik Cathedral \cite{McGuire2017Data}. For the approximation $I'$ of $I$ that the network should learn, we chose images rendered for the same frames at full resolution ($I$) and at four different reduced shading rates ($I'$). 

Table \ref{tab:validation} shows the measured statistics per scene for predicting each metric between reference images and their reduced versions on each scene's test set. Its consistent high accuracy, high coefficient of determination, and low variance in each scene's test set indicate that the network generalizes rather well: the model is capable of explaining most of the variance in each metric (high R$^2$) without over-fitting to specific scenes or states (visual examples of predictions are provided in Figure \ref{fig:metric_scenes}). 
We did not find a direct correlation between triangle/material count and the network's ability to predict perceptual metrics. In fact, the highest prediction accuracy was achieved on the most demanding scene in terms of geometry and the number of unique materials, 
Emerald Square at dusk, despite the network having only been trained on daylight scenes. The lowest scores were obtained in Bistro (Interior), which can be explained by the large number of specular objects it contains: since light sources are not explicitly encoded in the input, the network struggles to produce accurate predictions in previously unseen regions with specular materials. To test this theory, we created two variations of this scene: one with highly specular chrome materials and one with flat checkerboard textures applied everywhere (see supplemental material). As expected, the prediction quality, as indicated by $R^2$, is lower for the completely specular scene (FLIP:~$71\%$, PSNR:~$76\%$, LPIPS:~$64\%$, JNYang:~$70\%$, JNFLIP:~$70\%$).
However, for the same scene with only flat, checkered textures, the opposite is true: prediction quality rises, conversely, bringing it closer to the other scenes.
The network does not require a large number of training samples to achieve generalization: in our experiments, we found a negligible decrease in test accuracy---$0.04\%$---when an environment is not included as part of the training and found no benefit in using more than $500 - 2000$ captured frames on any environment (the exact number depends on the scene size). 

Finally, we recorded the run time for prediction and compared it against reference implementations of the corresponding metrics on the CPU and on the GPU (Python+PyTorch).
For our neural network, timings are independent of the metric it was trained on since it does not influence its architecture.
Inference with our network took $0.58s$/$2ms$ on CPU/GPU, respectively.
It is thus significantly faster than explicitly computing FLIP ($2.46s$/$190ms$ $\rightarrow$ $4.24\times$/$95.9\times$) and LPIPS ($13.6s$/$16.4ms$ $\rightarrow$ $23.5\times$/$8.2\times$).
For the much simpler PSNR, our approach is between $2\times$ and $10\times$ slower.

\subsection{Content-Adaptive Shading Application}
To assess a real-time use case, we implemented content-adaptive deferred shading in Falcor \cite{falcor} using our network, trained on JNYang and running on $16\times16$ tiles at $1080p$ resolution. We load the network into TensorRT and provide it with GBuffer-texture inputs in Falcor directly. For comprehensive results, we ran performance evaluation on five scenes that exhibit varying complexity in terms of geometry and materials: Suntemple, Bistro (Exterior), and the regular/specular/checkered Bistro (Interior). 
Frame times of our approach was compared with full-rate shading and a state-of-the-art VRS method \cite{yang2019adaptive}.
We considered two types of camera motion between frames: slow (resulting in 14\% previously unseen pixels per frame on average), and fast (31\% unseen on average), and evaluated 15 corresponding viewpoint pairs per scene and speed.

Inference with TensorRT requires a constant $\approx2.3$ ms per frame. For our approach to provide a benefit, it must amortize this overhead, which can only occur under appreciable fragment shader load. To simulate a pipeline comparable to interactive graphics applications (e.g., AAA video game titles), we created a synthetic load (50:1 arithmetic to memory) in the deferred fragment shader to bound full-rate shading performance to ~60 FPS.
In combination with our network's prediction, GPU hardware support for VRS yields a considerable performance gain across the board. 
For a slow-/fast-moving camera between frames, we achieved a
$1.12$/$1.14\times$ speedup for Suntemple, $1.17$/$1.18\times$ for Bistro (Exterior), and $1.42$/$1.41\times$ for the regular Bistro (Interior). The purely specular and checkered versions of the latter performed slightly better ($1.5$/$1.54\times$ and $1.48$/$1.52\times$, respectively): in both cases, this can be explained by the reduction of sharp features and high-frequency visual details in the scene, which enables the network to choose lower shading rates. 
In summary, VRS using our network reduced average frame times by at least $10\%$ compared to full-rate shading in all examined scenarios. The relative performance gain is boosted by the reduction of high-frequency features, permitting the use of lower shading rates.

For comparison with \citeauthor{yang2019adaptive} \shortcite{yang2019adaptive}, we used the same setup and configured the synthetic load so to have their approach match the target frame rate.
Since their base overhead is significantly lower than our network's inference time, our method trails behind \citeauthor{yang2019adaptive}'s at 60 FPS with slow camera motion on static scenes ($52.1$ FPS on average across all scenes $\rightarrow$ $0.87\times$). For fast camera motion, however, our method performs better ($1.03\times$) due to its ability to predict and use lower shading rates in unseen regions, rather than defaulting to full resolution.
Using an even heavier load (30 FPS target), our method prevails as soon as camera motion occurs ($1.11\times$ at slow, and $2.16\times$ at fast motion). Hence, even given the early state of dedicated GPU inferencing hardware, our learning-based approach can provide clear benefits in such demanding scenarios.

\subsection{Limitations}
\begin{figure}
  \centering
  \begin{subfigure}{0.23\textwidth}
      \includegraphics[width=\textwidth]{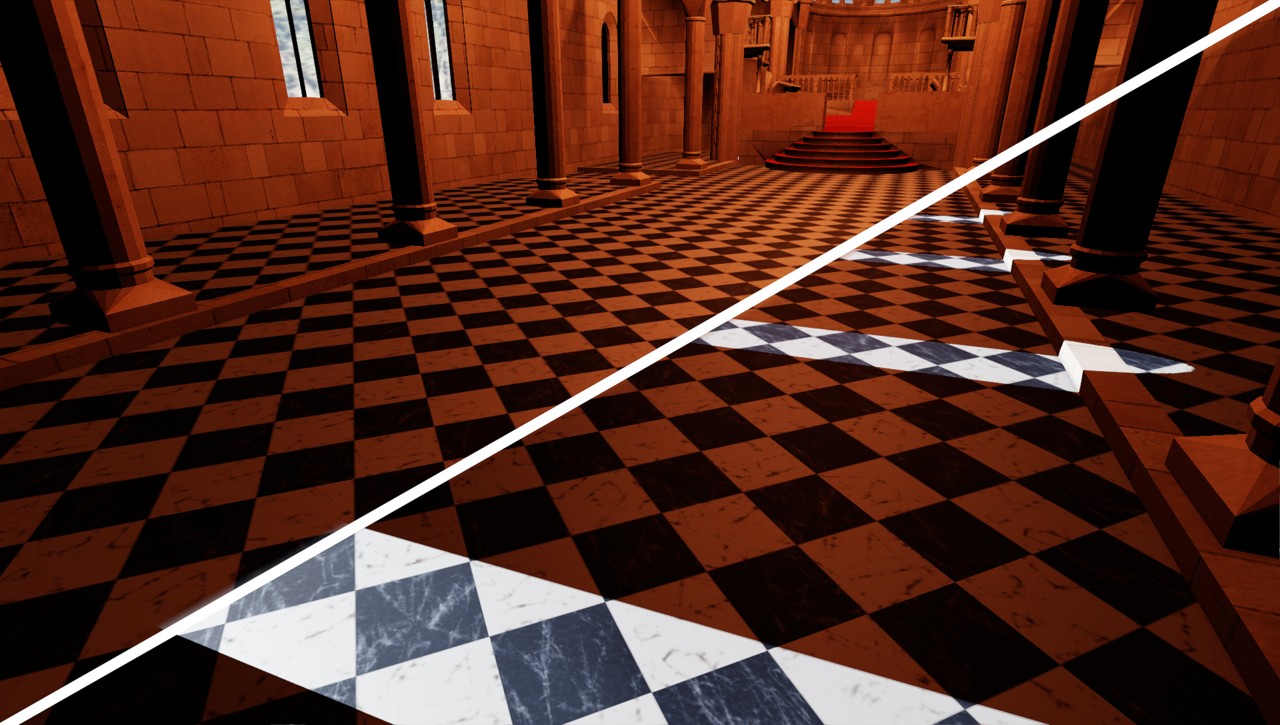}
      \raisebox{13mm}[0mm][0mm]{\footnotesize
      \hspace{1mm} \color{white} frame $N-1$}
      \raisebox{11mm}[0mm][0mm]{\footnotesize \color{white} \hspace{27mm} frame $N$}
      \vspace{-8mm}
      \caption{Light flash through windows}
  \end{subfigure}
  \hfill
  \begin{subfigure}{0.23\textwidth}
      \includegraphics[width=\textwidth]{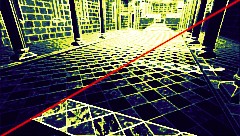}
      \raisebox{13mm}[0mm][0mm]{\footnotesize
      \hspace{1mm} \color{white} frame $N$}
      \raisebox{11mm}[0mm][0mm]{\footnotesize \color{white} \hspace{27mm} frame $N+1$}
      \vspace{-8mm}
      \caption{Delayed predicted metric}
  \end{subfigure}
  \begin{subfigure}{0.23\textwidth}
      \includegraphics[width=\textwidth]{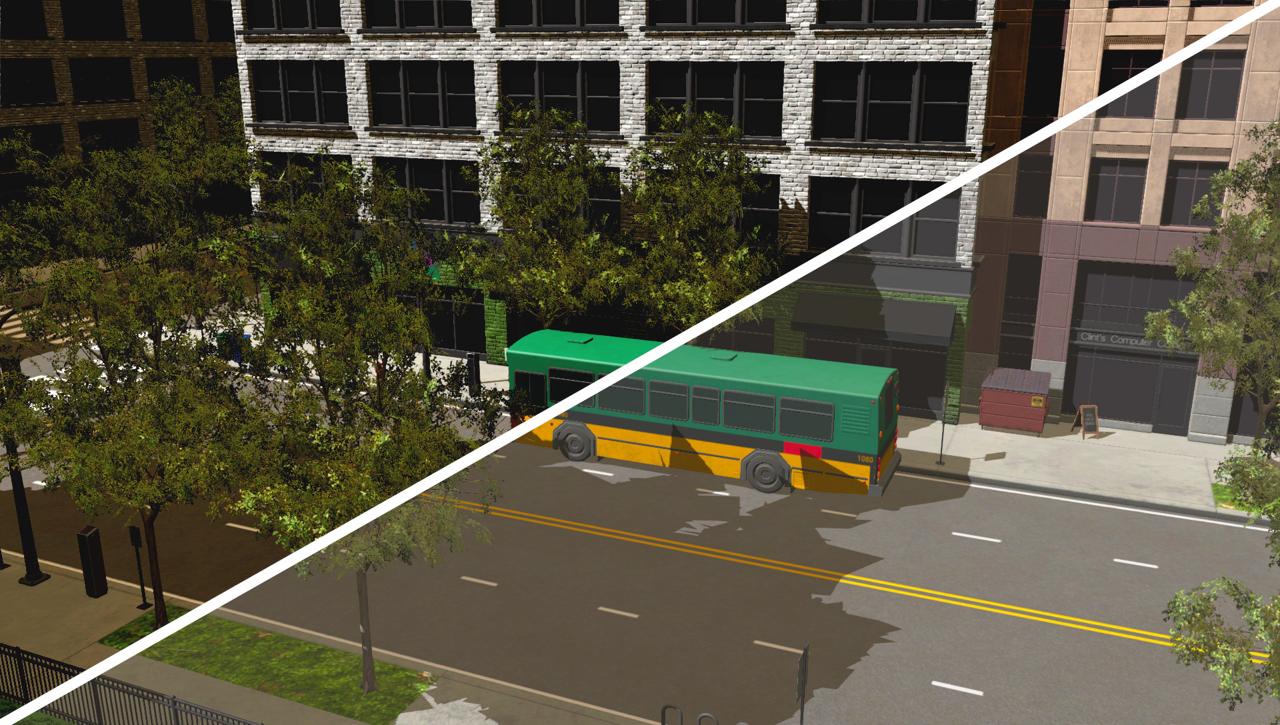}
      \raisebox{13mm}[0mm][0mm]{\footnotesize
      \hspace{1mm} \color{white}  frame $N-1$}
      \raisebox{7mm}[0mm][0mm]{\footnotesize \color{white} \hspace{0mm} frame $N$}
      \vspace{-4mm}
      \caption{Tone mapper ACES/Reinhardt}
  \end{subfigure}
  \hfill
  \begin{subfigure}{0.23\textwidth}
      \includegraphics[width=\textwidth]{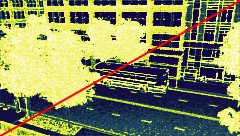}
      \raisebox{13mm}[0mm][0mm]{\footnotesize
      \hspace{1mm} frame $N-1$}
      \raisebox{7mm}[0mm][0mm]{\footnotesize \color{white} \hspace{0mm} frame $N$}
      \vspace{-4mm}
      \caption{Unchanged prediction}
  \end{subfigure}
  \begin{subfigure}{0.23\textwidth}
    \includegraphics[width=\textwidth]{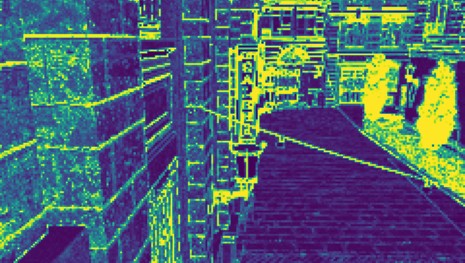}
    \caption{Ground truth}
  \end{subfigure}
  \hfill
  \begin{subfigure}{0.23\textwidth}
      \includegraphics[width=\textwidth]{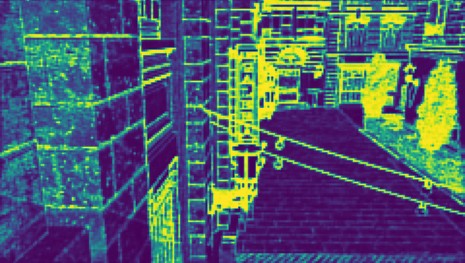}
      \caption{Predicted}
  \end{subfigure}
  
  \caption{(a,b) Reliance on reprojection can cause the network to react to sudden lighting changes in previously seen regions with a delay of one frame. (c,d) Changing tone-mapping method also does not result in immediate different predictions. (e,f) Incorrect previous frame reprojections can cause our network to hallucinate duplicated objects due to surface information mismatch.}
  \label{fig:limitations}
\end{figure}

The key purpose of our approaches is to enable optimizations in real-time applications by predicting the---otherwise expensive---pixel shading result. This naturally impedes its ability to account for factors that are unknown prior to pixel shading.
We circumvent this issue by relying on reprojection and G-buffer data, the latter of which may not contain all information affecting the final color generation (e.g., light source position, cf.\ Figure \ref{fig:channels}). Hence, similar to other state-of-the-art methods \cite{yang2019adaptive}, the network is bound to make assumptions about such effects based on previously seen regions. If an effect cannot be predicted from G-buffer data alone, it may only react to it in the next frame, when its reprojection becomes available. This includes temporal inconsistencies in the scene (e.g., sudden disocclusion of a strong light source), reflections, and modifying of rendering settings or post-processing effects (Figure \ref{fig:limitations}).
However, in this paper, we have shown that our approach can be trained to discard reprojected color and substitute information derived from G-buffer data instead. Hence, it may be trained to adapt to sudden changes immediately. For instance, in the case of a disoccluded light source, this could be achieved by providing additional input tracking changes in the binary screen-space shadowing information between frames.
For more complex effects, like reflections or fog, more sophisticated solutions may be needed to provide suitable, inexpensive approximations of the required information to the network. The decision of trading a single-frame delay of predicted effects for larger input sets should then depend on the user's expected attention to them.
Future work may explore under which circumstance reprojection may be omitted and instead replaced by additional, equally expressive encodings or estimates of important scene features, such as light sources and reflections. Tackling this challenge would come with the advantage of providing a unified solution for both seen and unseen regions.

Although the achieved performance in real-time applications is acceptable with our approach, it incurs an overhead that limits its applicability. For slow-moving changes, selective reuse of predictions could significantly alleviate this issue, which we aim to pursue in future work.
In our proof-of-concept, the naive screen-space reprojection used is not precise, which can sometimes cause our network to hallucinate thin objects' duplicates due to material and reprojection data inconsistency (Figure \ref{fig:limitations}). This could be improved upon by using state-of-the-art, non-screen-space reprojection.
    \section{Conclusion}
In this paper, we have presented a method for training and predicting perceptual metrics using a learning-based approach. The proposed network architecture is compact enough to make predictions with high accuracy in real-time, without relying on a reference or rendered image. 
We have shown how to tackle common machine learning problems, such as unbalanced training data, with specialized solutions for our task that anticipate the eventual real-time applications.
Furthermore, we have shown how the concept of visually-based decision-making with just-noticeable differences can be directly integrated into the learning process.

Our solution can be used to predict various metrics and generalizes well to new scenes and applications.
By exploiting recent advances in GPU hardware, inference can be performed in real-time, thus opening the door for new uses of visual error metrics in real-time rendering applications. 
Our exemplary content-adaptive shading setup shows that, while direct execution of our network per-frame may not always be expedient, visually-based decision-making can already be performed at highly interactive frame rates.
Hence, applications with very demanding shading or only occasional prediction that is amortized over time are likely to benefit from our solution. Furthermore, it is safe to assume that future hardware generations will significantly improve upon neural network inference speed.
To enable experimentation and research of such applications, we have published our full codebase for capturing, learning, and applying relevant metrics to 3D scenes and used datasets at \url{jaliborc.github.io/rt-percept}.
    
    \bibliographystyle{ACM-Reference-Format}
    \bibliography{main}
    
    \clearpage \appendix
    \section{Data Capture Pipeline}
\begin{figure}
  \centering
  \begin{subfigure}{0.23\textwidth}
    \includegraphics[width=\textwidth]{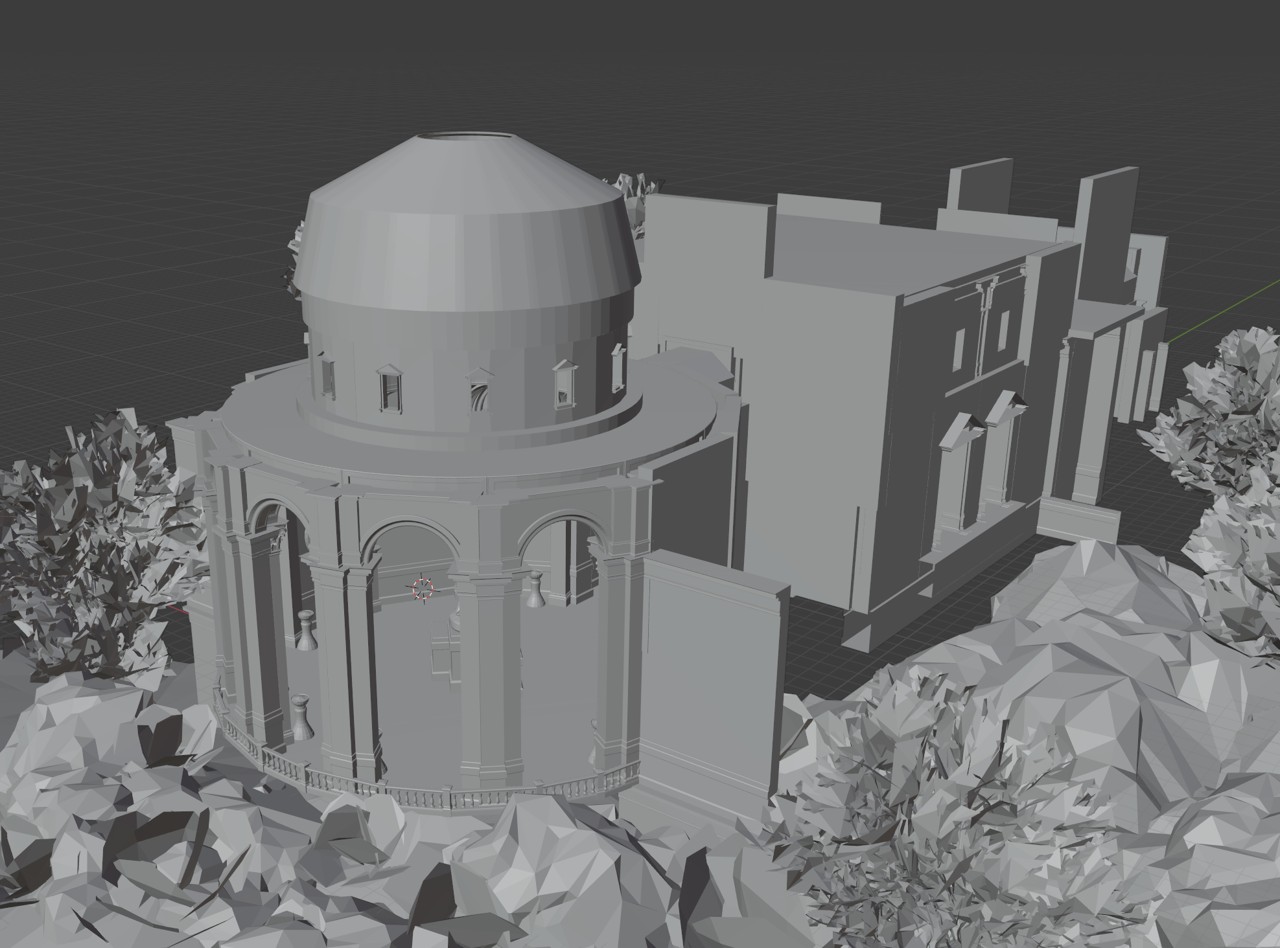}
    \caption{Suntemple Scene}
  \end{subfigure}
  \begin{subfigure}{0.23\textwidth}
    \includegraphics[width=\textwidth]{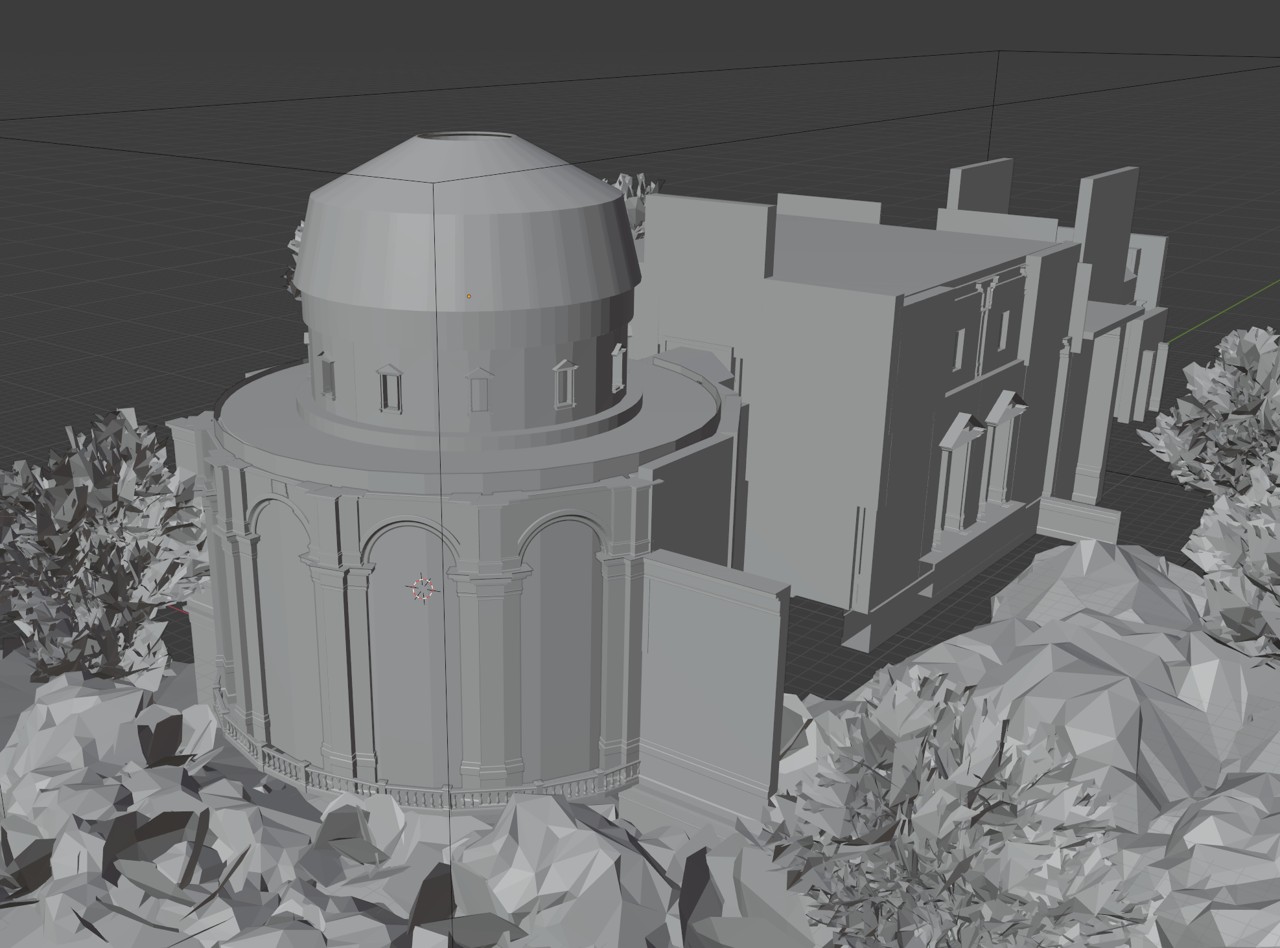}
    \caption{Defined Boundaries} \label{fig-blender-domain}
  \end{subfigure}
  \begin{subfigure}{0.23\textwidth}
    \includegraphics[width=\textwidth]{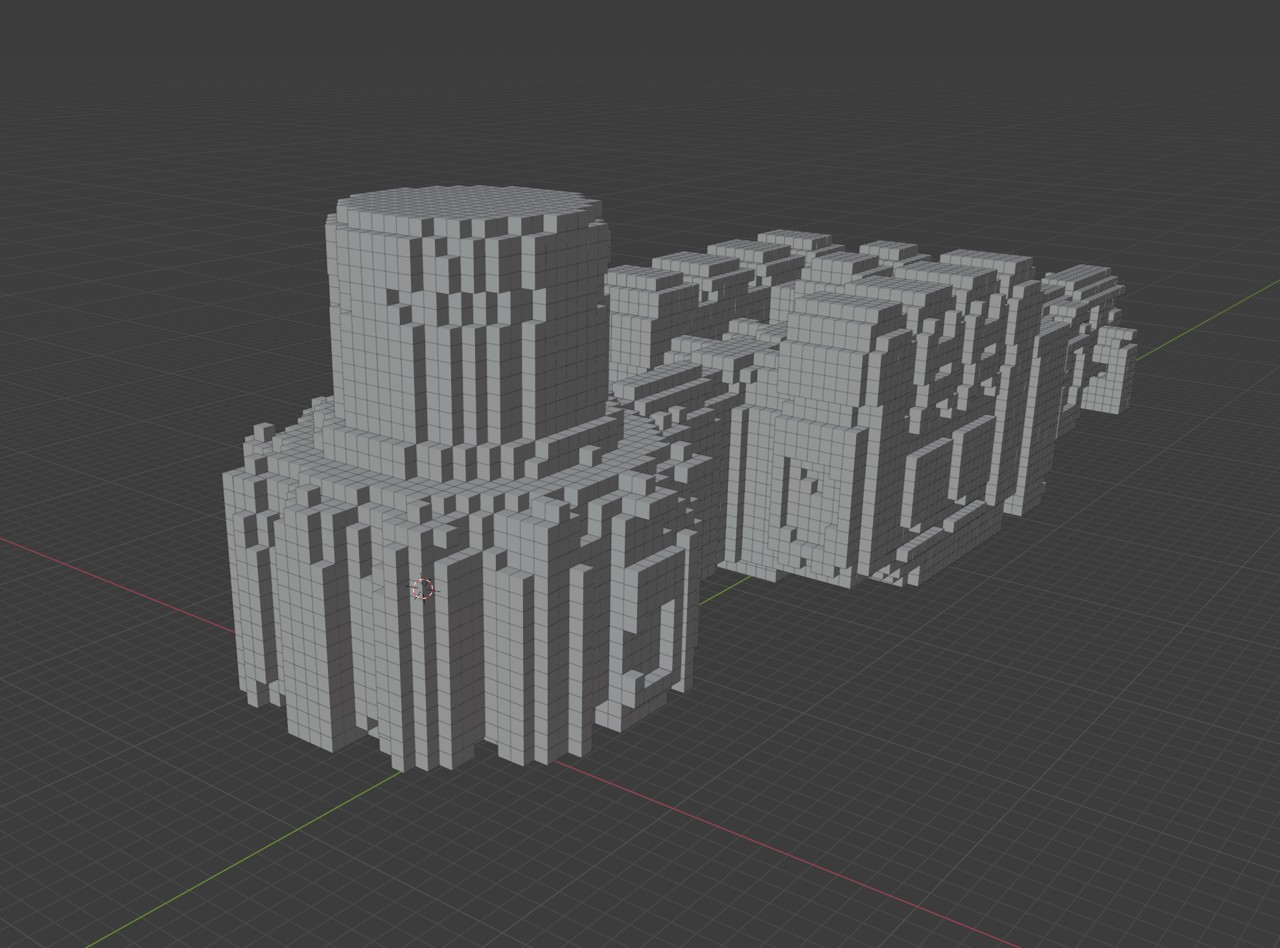}
    \caption{Volume Marching}
  \end{subfigure}
  \begin{subfigure}{0.23\textwidth}
    \includegraphics[width=\textwidth]{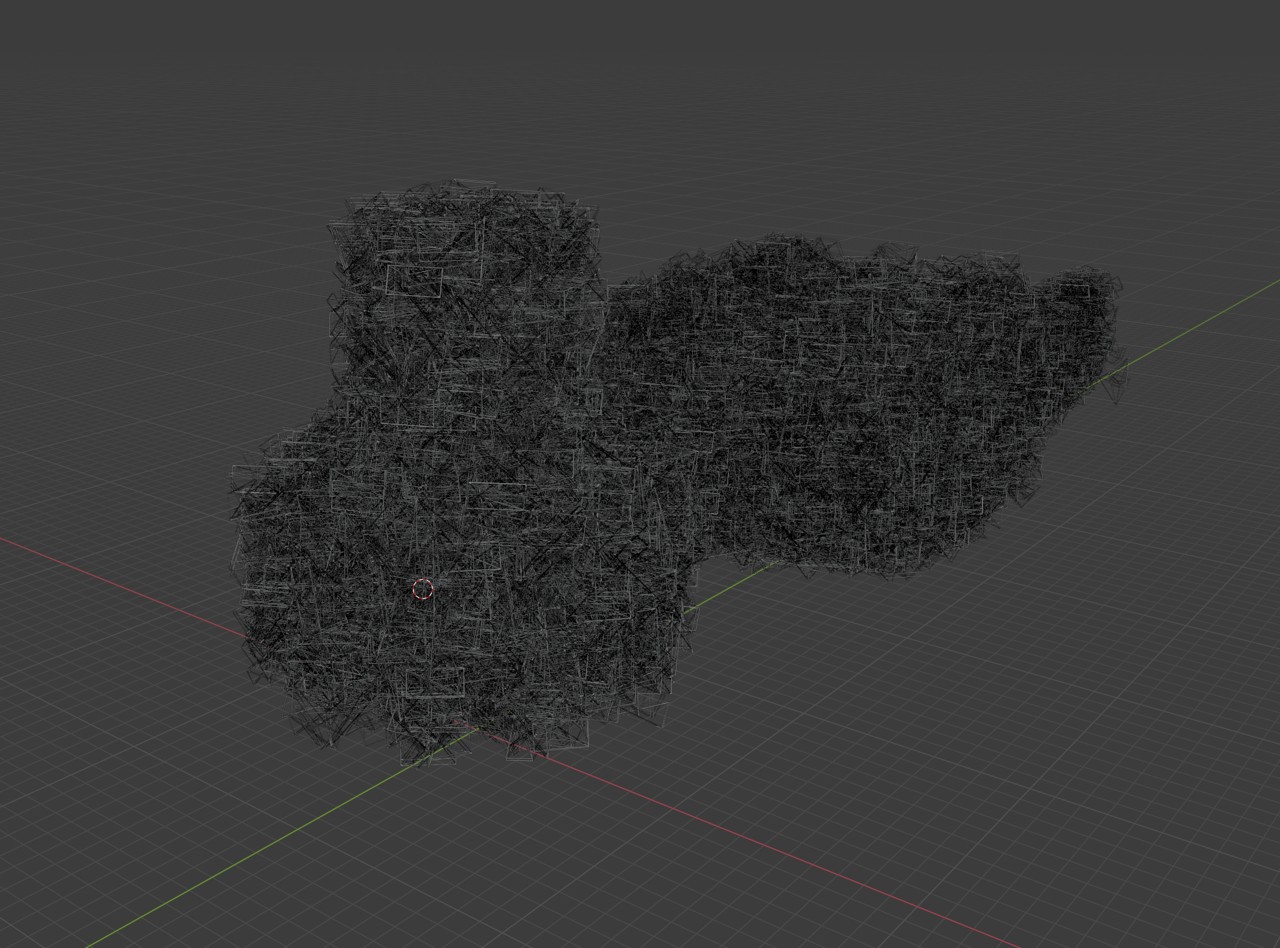}
    \caption{Chosen Viewpoints}
  \end{subfigure}
  \caption{The different stages in our Blender data capture pipeline. (b) is the only manual step required on open maps.}
  \label{fig-blender}
\end{figure}

To avoid bias in our training data and ensure consistency in dataset generation during the development, we created a pipeline to automate the generation of our datasets. It was implemented as a combination of a Blender \shortcite{blender} script tool and custom renderers on Falcor \cite{falcor}. See Figure \ref{fig-blender} for an overview of its stages.

For any given scene, we define a cube in the scene centered on a valid viewpoint. Then, we perform flood fill using the cube geometry to create a voxel domain for potentially valid viewpoints. Flooding alone works in closed environments but would leak on any open environment to an infinite domain. To solve this, we take inspiration from game-level design and manually add invisible walls to the scene to limit the valid voxel domain (see Figure \ref{fig-blender-domain}). Finally, we select random 3D points in the voxel space paired with random 3D directions and test whether they would make valid viewpoints.

Different criteria could be used to validate viewpoints. We filter them based on two:

\begin{enumerate}
    \item Whether a template camera geometry placed on the viewpoint position and direction intersects with the scene.
    \item Whether it renders a minimum amount of visible geometry, measured in percentage of rendered pixels ($80\%$).
\end{enumerate}

The first requirement prevents the dataset from containing viewpoints where the camera intersects with the scene geometry, while the latter avoids an exterior dataset being filled with viewpoints looking at the skybox, for example.

Having a predefined amount of valid viewpoints selected, we randomly select for each a corresponding ``previous viewpoint". This simulates the environment being explored by the player and is necessary to generate reprojection training data. To do so, we select random valid viewpoints just as before, but each in very close proximity to its corresponding ``next frame". Besides the two aforementioned criteria, we also check whether a raycast from one viewpoint to the next intersects with geometry to verify there is an open path between the two.

Finally, for each previous/next viewpoint pair entry, we render the set of g-buffers, reprojected color, and final renders at different shading rates in Falcor.

\section{Hyperparameter Details}
\begin{table}
  \caption{Network layers dynamic parameters for $w=16$.}
  \label{tab:layers}
  \begin{tabular}{ccccc}
    \toprule
    Layer & In Channels & Out Channels & Groups & Pooling Factor\\
    \midrule
    $1$ & \# input data & 16 & $1$ & $1$\\
    $2$ & $16$ & $16$ & $1$ & $2$\\
    $3$ & $16$ & $16$ & $4$ & $2$\\
    $4$ & $16$ & $16$ & $8$ & $2$\\
    $5$ & $16$ & \# predictions & $1$ & $2$\\
  \bottomrule
\end{tabular}
\end{table}

In this Section, we elaborate on hyperparameter details we found less crucial to include in Section \ref{sec:network} of the paper. As mentioned in that Section, all our convolutions work on a $3\times3$ kernel. They also have a stride, padding, and dilation equal to $1$ to ensure the image output size of the convolution matches the input size. We chose a latent channel dimension of $16$ as our testing showed us that it is the lowest one can pick before there is an evident loss of prediction quality. However, one can go as low as $8$ latent channels before the network becomes unusable. As an illustrative example, Table \ref{tab:layers} shows how grouping and max-pooling is used throughout the network five layers when the intended output tiling size is $16\times16$ pixels ($w=16$).

Initially, we performed training using stochastic gradient descent
However, it required a very low learning rate of $10^{-6}$ with an additional exponential learning decay of $1-10^{-4}$ to remain stable. We recommend using root mean squared propagation, which allows for a much faster training with a learning rate of $10^{-4}$, although it still requires the same level of decay.

\section{Specialization and Distribution}

Clearly, the effectiveness of our approach is dictated by the data that is provided to it during training. While we found that it is capable of generalizing across various scenes and settings, it inherently specializes to the shading models it observes: a network trained on GGX cannot reason about the errors observed with, e.g., artistic toon shaders. However, due to its compact design, the full network we used in our VRS use case requires only 22 KiB of storage. Hence, in practice, it would be entirely viable to prepare (and ship) several instances of trained networks for, e.g., differently stylized animations, cut scenes or even level designs.

\section{Additional Illustrations}
In this Section, we present additional visuals to help illustrate selected properties of our approach. We include specific details of our transforms, a side-by-side comparison of the metrics we tested in this paper shown on the same scene, assessed variations of Bistro (Interior), additional examples and comparisons of VRS using our network with different metrics, and hallucinations that can sometimes occur in our network.

\begin{figure*}
  \centering
  \includegraphics[width=\textwidth]{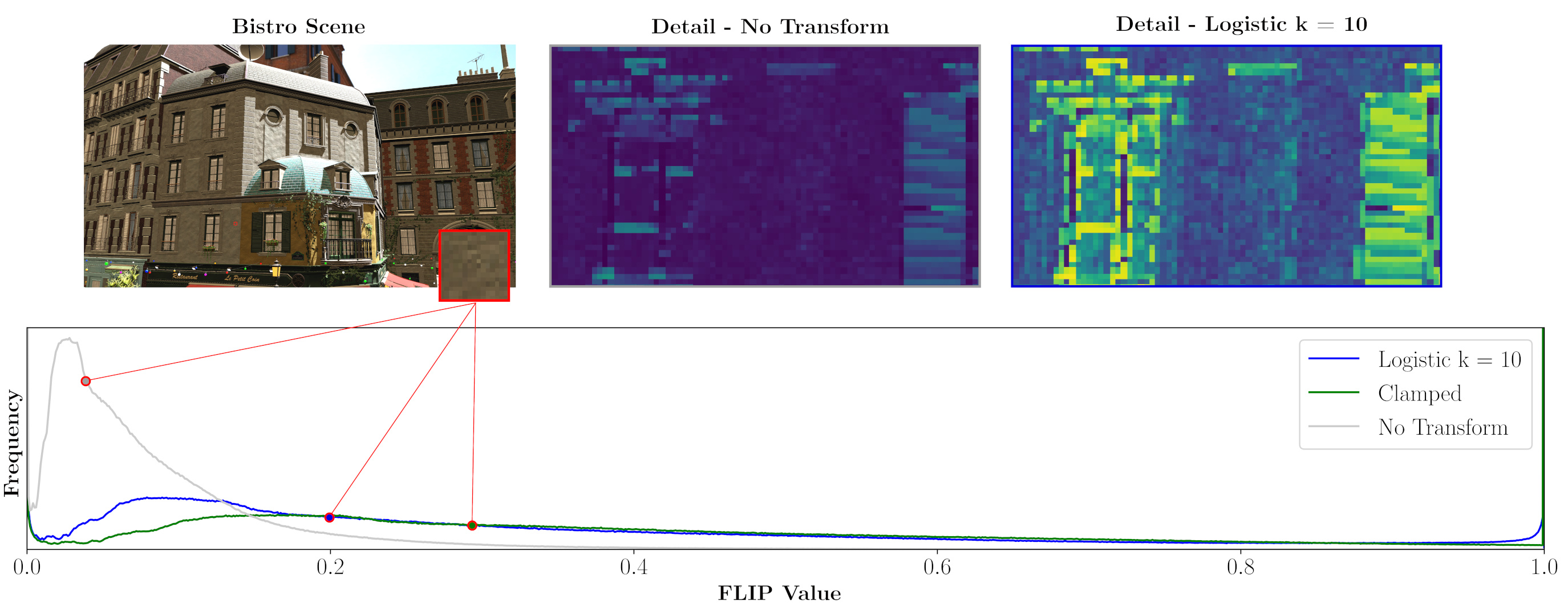}
  \caption{Visualization of region in the training set distribution, according to different FLIP parameter spaces. The clamping transform approaches a uniform distribution, at the cost of ignoring differences between the highest values (notice the spike at $\hat{Y} = 1.0$). The logistic transform achieves a similar result but gives greater importance to nuanced decisions. Both methods preserve all available information.}
  \label{fig:transform-dist}
\end{figure*}

\begin{figure*}
    \centering
    \begin{tikzpicture}[scale=0.75]
        \begin{axis}[
            width=\textwidth,
            height=8cm,
            xlabel=Absolute Value,
            ylabel=Adapted Value,
            legend pos=south east
        ]
            \addplot[color=gray,line width=1pt,domain=0:1]{x};
            \addlegendentry{No Transform}
            \addplot[color=cyan,line width=1pt,domain=0:1]{((1 / (1 + e ^ ((-x+0.25)*2))) - (1 / (1 + e ^ ((+0.25) * 2)))) / ((1 / (1 + e ^ ((-1+0.25) * 2))) - (1 / (1 + e ^ ((+0.25)*2))))};
            \addlegendentry{Logistic $k=2$}
            \addplot[color=blue,line width=1pt,domain=0:1]{((1 / (1 + e ^ ((-x+0.25)*5))) - (1 / (1 + e ^ ((+0.25) * 5)))) / ((1 / (1 + e ^ ((-1+0.25) * 5))) - (1 / (1 + e ^ ((+0.25)*5))))};
            \addlegendentry{Logistic $k=5$}
            \addplot[color=teal,line width=1pt,domain=0:1]{((1 / (1 + e ^ ((-x+0.25)*10))) - (1 / (1 + e ^ ((+0.25) * 10)))) / ((1 / (1 + e ^ ((-1+0.25) * 10))) - (1 / (1 + e ^ ((+0.25)*10))))};
            \addlegendentry{Logistic $k=10$}
            \addplot[color=green,line width=1pt,domain=0:1]{min(2*x, 1)};
            \addlegendentry{Clamped}
        \end{axis}
    \end{tikzpicture}
    \caption{Visualization of the parameter spaces created as a result of our transforms, for a training distribution average value $\mu_{Y} = 0.25$.}
    \label{fig:transform-graph}
\end{figure*}
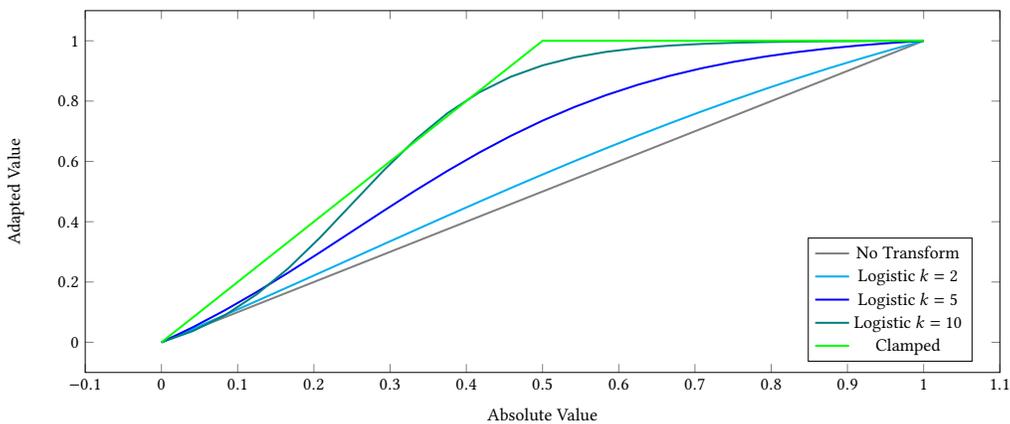

\begin{figure*}
  \centering
  \begin{subfigure}{0.33\textwidth}
    \includegraphics[width=\textwidth]{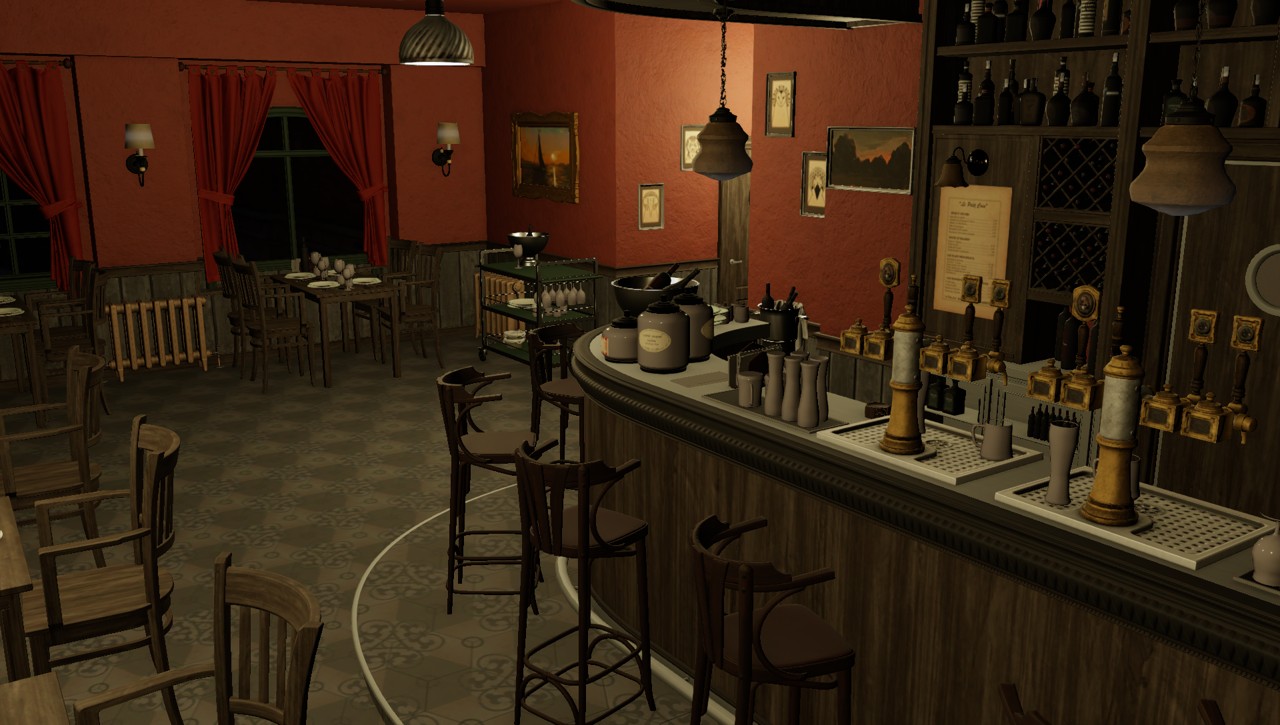}
    \caption{Bistro scene}
  \end{subfigure}
  \begin{subfigure}{0.33\textwidth}
    \includegraphics[width=\textwidth]{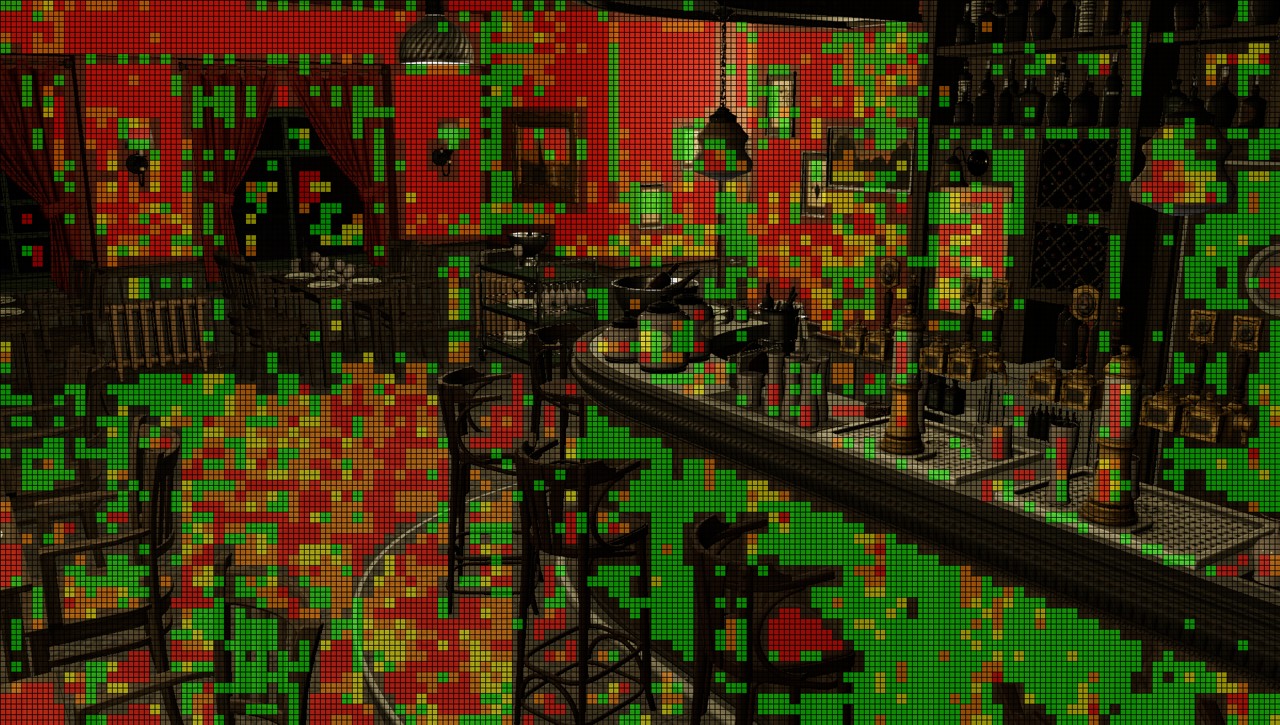}
    \caption{Ground truth}
  \end{subfigure}
  \begin{subfigure}{0.33\textwidth}
    \includegraphics[width=\textwidth]{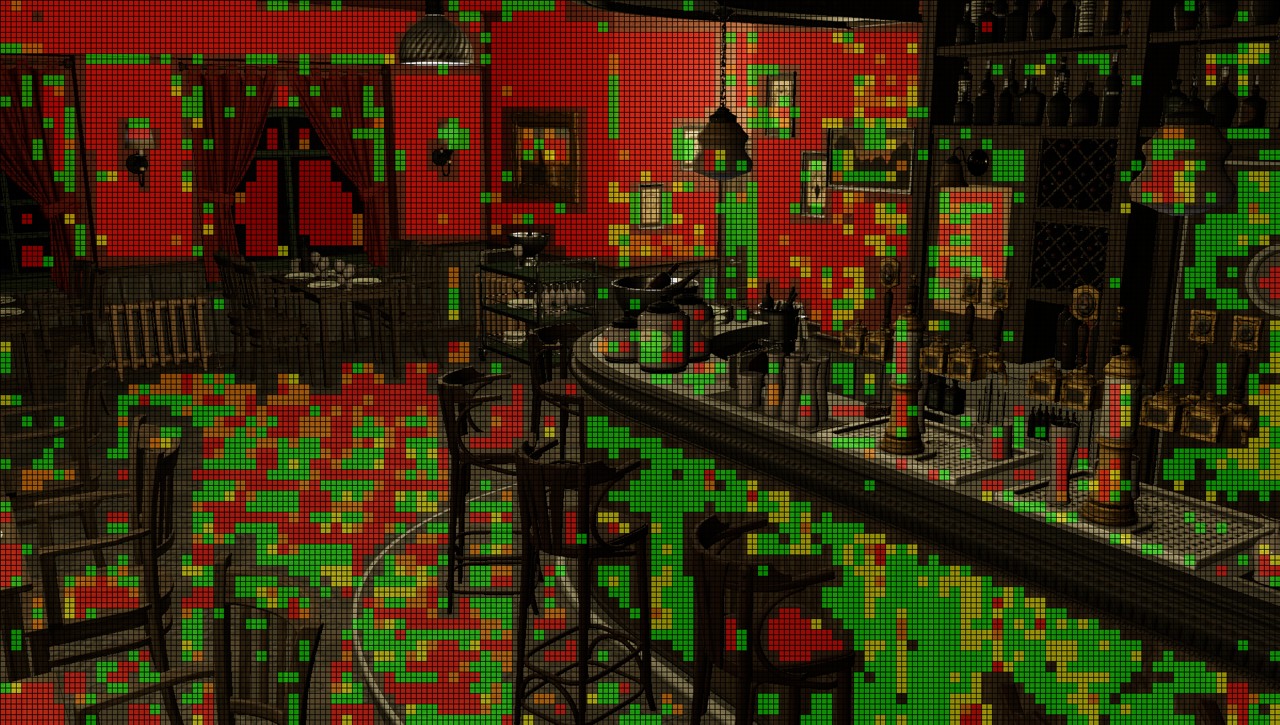}
    \caption{Predicted}
  \end{subfigure}
  \caption{Example of variable rate shading on 16x16 tiles using offline ground truth JNFLIP vs realtime network JNFLIP prediction.}
  \label{fig:gt-vs-prediction}
\end{figure*}

\begin{figure*}
  \centering
  \begin{subfigure}{0.33\textwidth}
    \includegraphics[width=\textwidth]{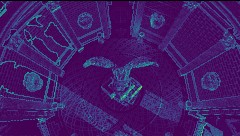}
    \caption{Yang}
  \end{subfigure}
  \hfill
  \begin{subfigure}{0.33\textwidth}
    \includegraphics[width=\textwidth]{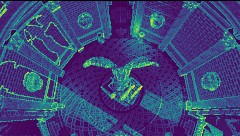}
    \caption{FLIP}
  \end{subfigure}
  \hfill
  \begin{subfigure}{0.33\textwidth}
    \includegraphics[width=\textwidth]{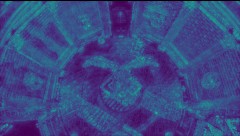}
    \caption{LPIPS}
  \end{subfigure}
  \vspace{.5em}
  \hfill
  \begin{subfigure}{0.33\textwidth}
    \includegraphics[width=\textwidth]{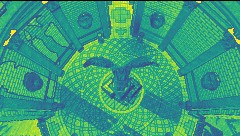}
    \caption{PSNR}
  \end{subfigure}
  \hfill
  \begin{subfigure}{0.33\textwidth}
    \includegraphics[width=\textwidth]{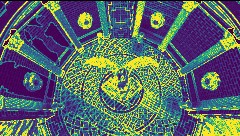}
    \caption{JNYang}
  \end{subfigure}
  \hfill
  \begin{subfigure}{0.33\textwidth}
    \includegraphics[width=\textwidth]{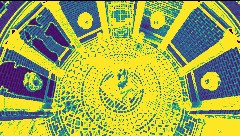}
    \caption{JNFLIP}
  \end{subfigure}
  \caption{Ground-truth side by side comparison of the metrics mentioned in this work for the same frame when considering half-resolution rendering.}
  \label{fig:metric-comparison}
\end{figure*}

\begin{figure*}
  \centering
  \begin{subfigure}{0.39\textwidth}
    \centering
    \includegraphics[height=3.9cm]{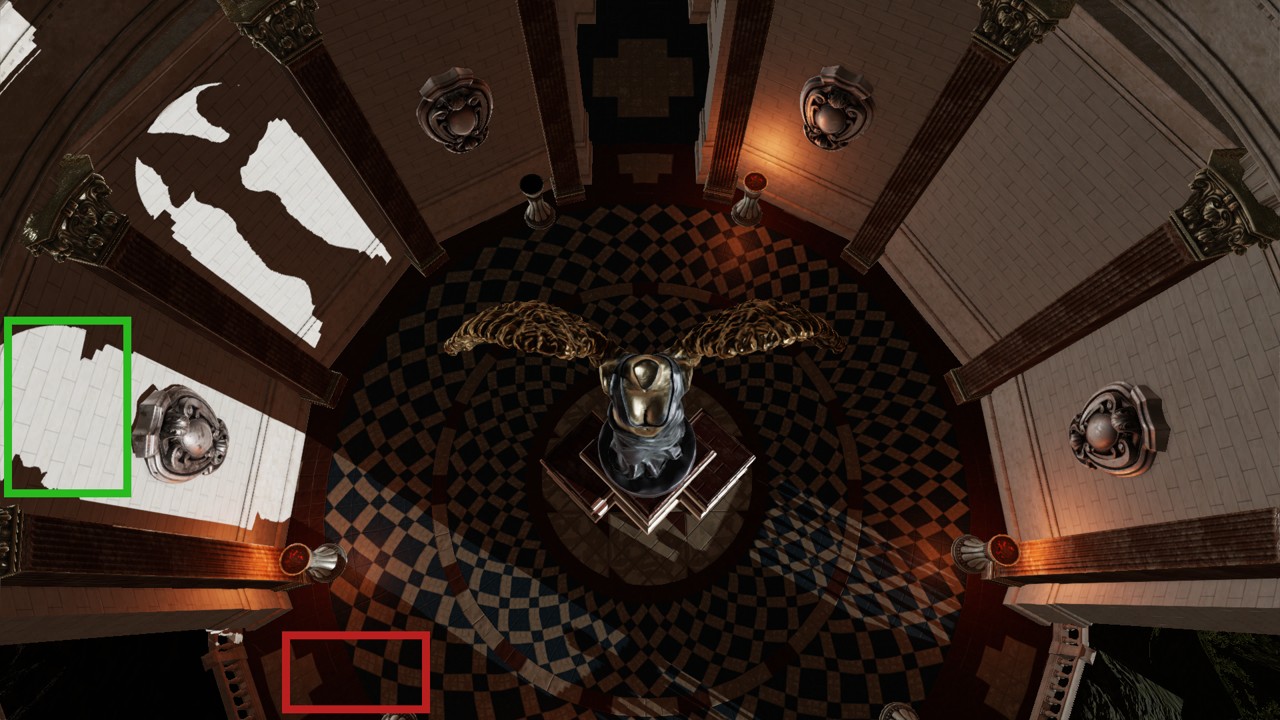}
    \caption{Suntemple scene with bright/dark areas}
  \end{subfigure}
  \begin{subfigure}{0.44\textwidth}
    \centering
    \includegraphics[height=1.95cm]{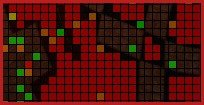}
    \includegraphics[height=1.95cm]{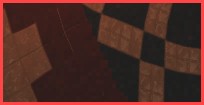}\\
    \includegraphics[height=1.95cm]{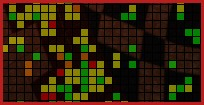}
    \includegraphics[height=1.95cm]{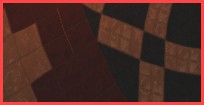}
    \caption{VRS based on predicted FLIP (top) vs JNFLIP (bottom)}
  \end{subfigure}
  \begin{subfigure}{0.158\textwidth}
    \centering
    \includegraphics[height=1.95cm]{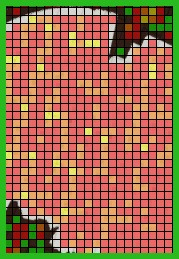}
    \includegraphics[height=1.95cm]{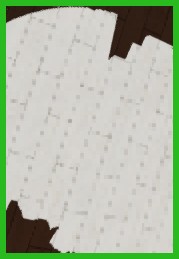}\\
    \includegraphics[height=1.95cm]{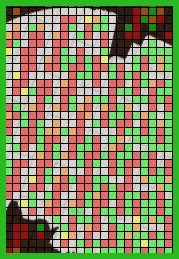}
    \includegraphics[height=1.95cm]{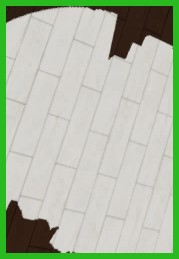}
    \caption{Yang et al. vs JNFLIP}
  \end{subfigure}
  \caption{JNFLIP can provide visual benefits in content-adaptive shading. FLIP is not normalized for local brightness and thus underestimates dark regions. The method by Yang \citeauthor{yang2019adaptive} \shortcite{yang2019adaptive} can struggle in shiny or overly exposed regions. JNFLIP handles both cases gracefully.}
  \label{fig:jnflip}
\end{figure*}

\begin{figure*}
  \centering
  \begin{subfigure}{0.33\textwidth}
    \centering
    \includegraphics[width=\textwidth]{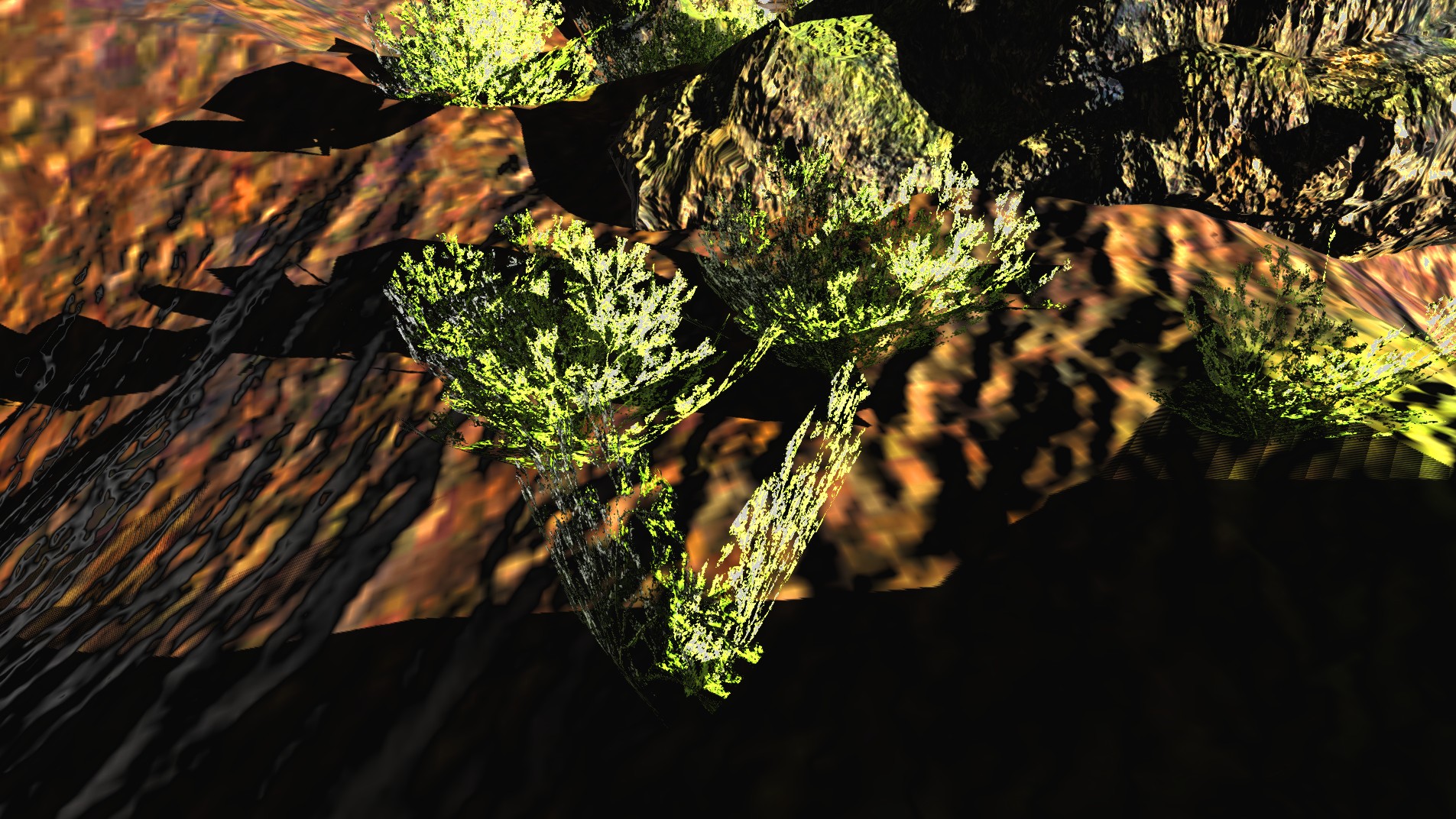}
    \caption{Suntemple, exterior}
  \end{subfigure}
  \begin{subfigure}{0.33\textwidth}
    \hfill
    \centering
    \includegraphics[width=\textwidth]{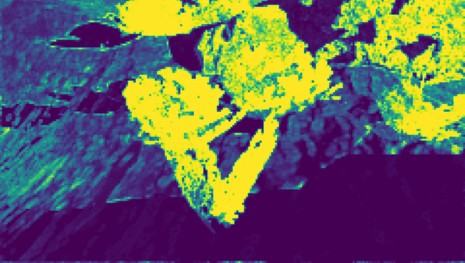}
    \caption{Ground truth FLIP}
  \end{subfigure}
  \begin{subfigure}{0.33\textwidth}
    \hfill
    \centering
    \includegraphics[width=\textwidth]{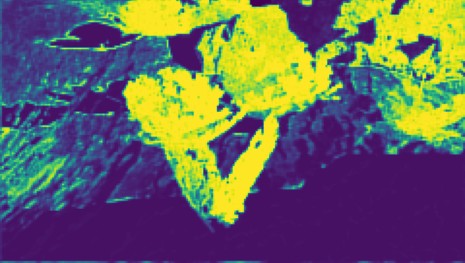}
    \caption{Predicted FLIP}
  \end{subfigure}
  \caption{A network trained strictly indoors in the Suntemple scene produces accurate FLIP predictions for bushes and rocks on the exterior.}
  \label{fig:temple-outdoors}
\end{figure*}

\begin{figure*}
  \centering
  \begin{subfigure}{0.49\textwidth}
    \centering
    \includegraphics[width=\textwidth]{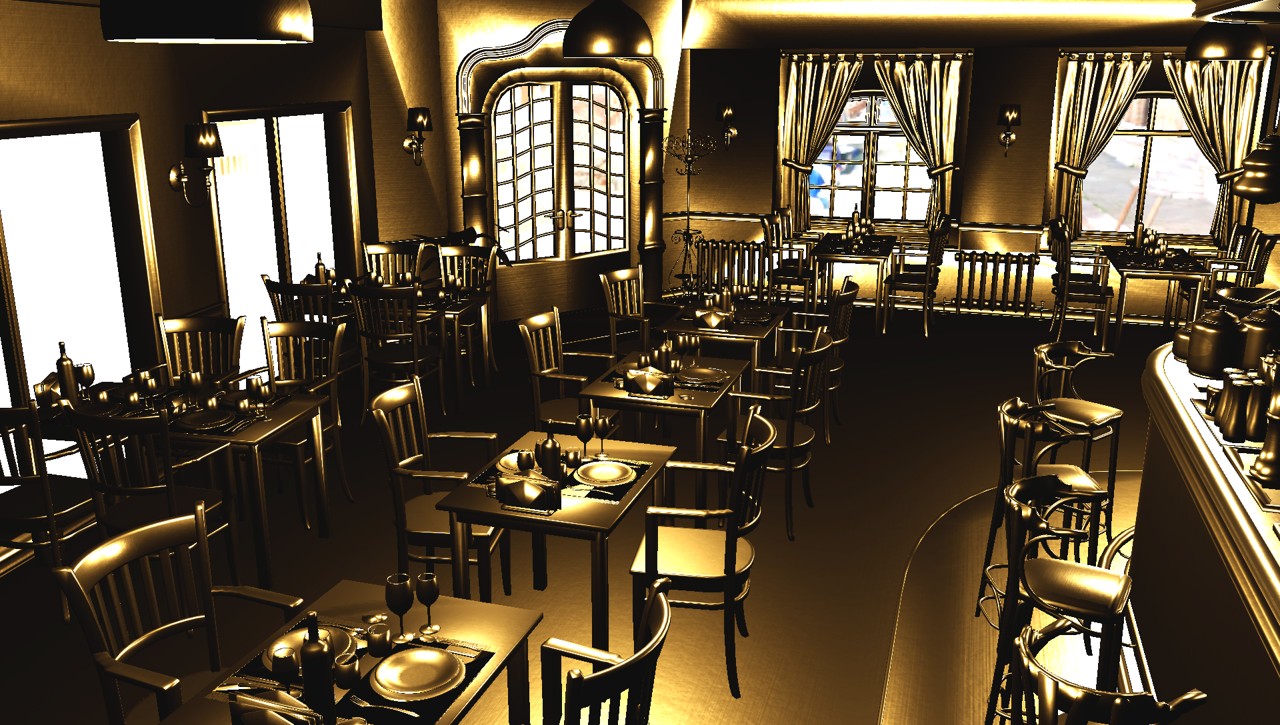}
    \caption{Bistro, interior, highly specular}
  \end{subfigure}
  \hfill
    \begin{subfigure}{0.49\textwidth}
    \centering
    \includegraphics[width=\textwidth]{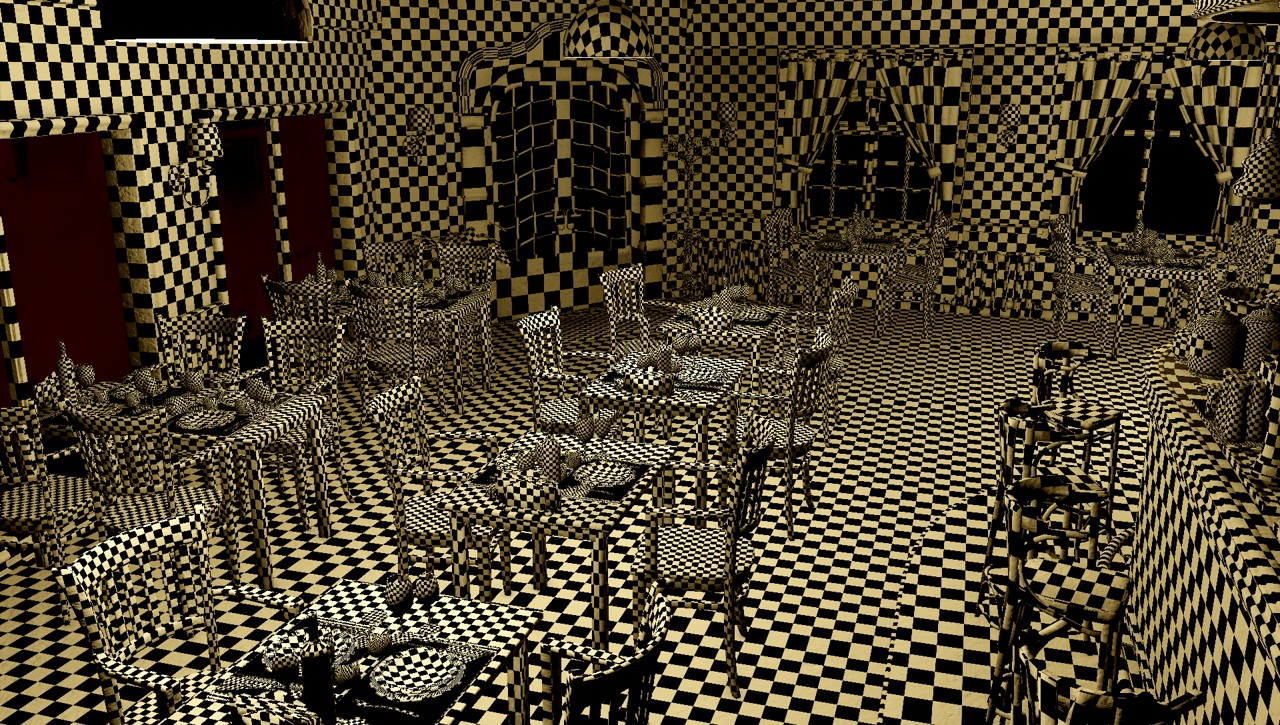}
    \caption{Bistro, interior, high-frequency textures}
  \end{subfigure}
  \caption{Two modified scenes used for evaluating the influence of (a) highly specular materials and (b) simple checkerboard textures on the prediction quality and performance of our approach.}
  \label{fig:spec-hf}
\end{figure*}

\begin{figure*}
  \centering
  \begin{subfigure}{0.33\textwidth}
    \includegraphics[width=\textwidth]{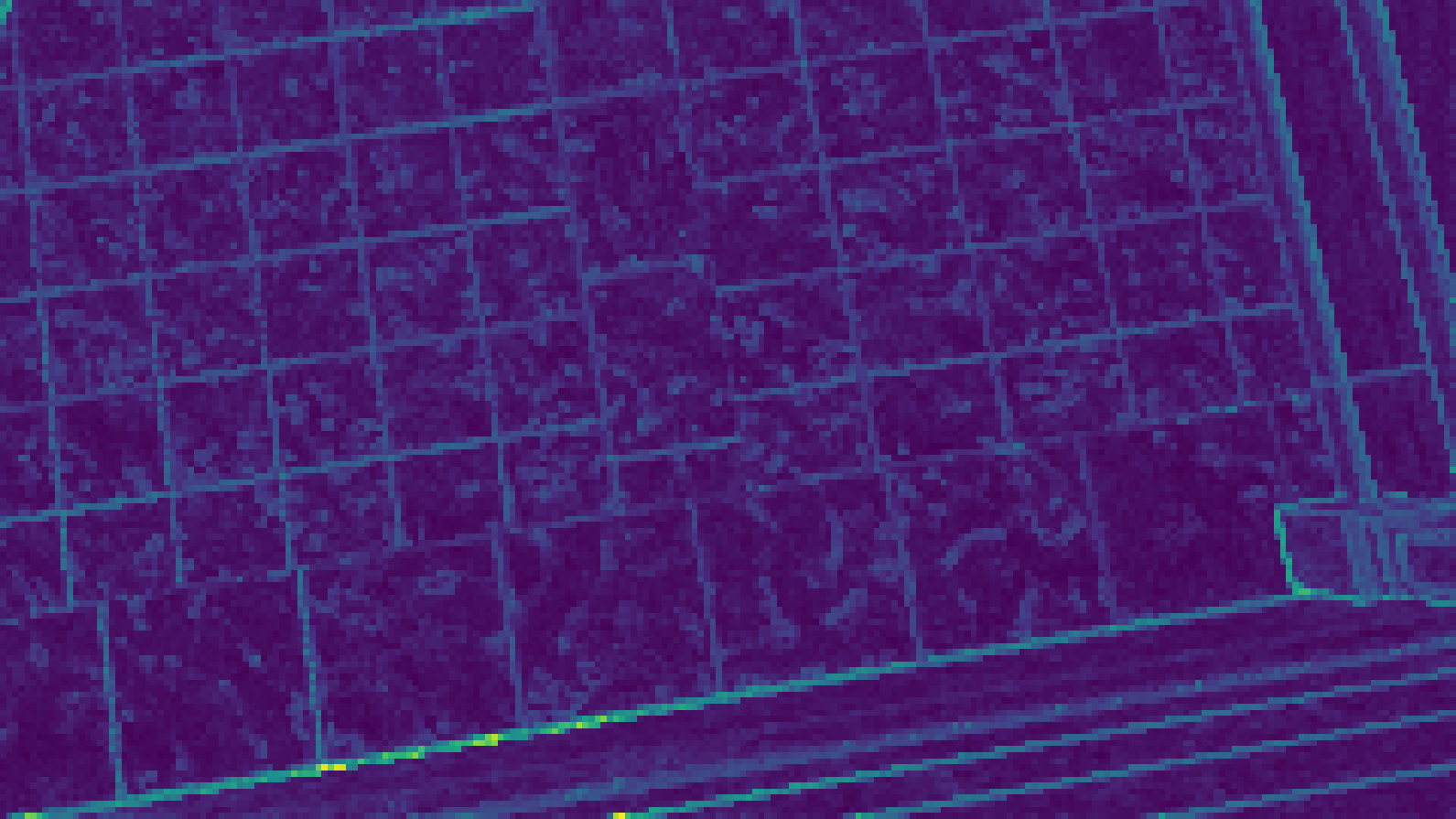}
    \caption{2x2}
  \end{subfigure}
  \hfill
  \begin{subfigure}{0.33\textwidth}
    \includegraphics[width=\textwidth]{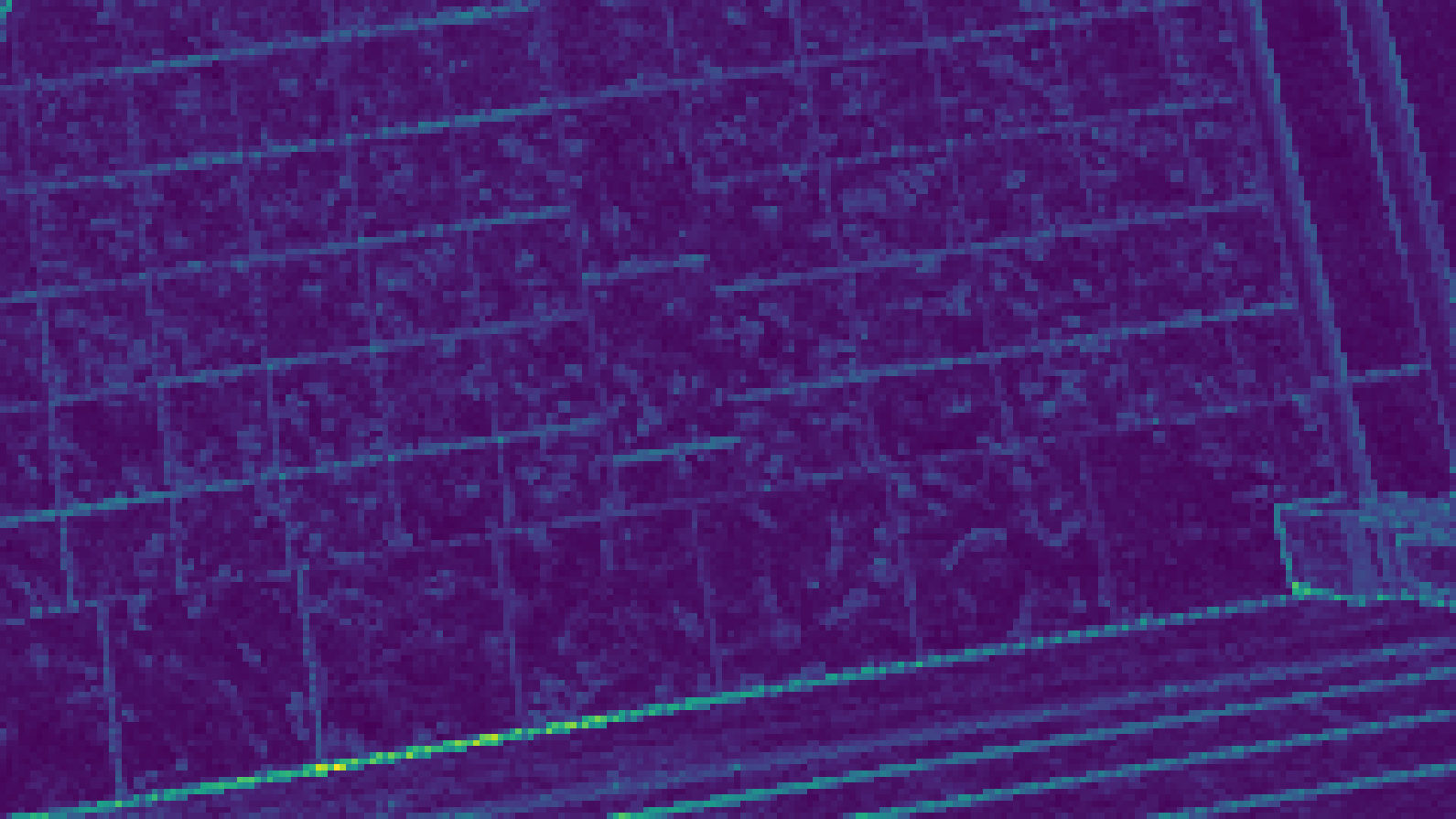}
    \caption{4x2}
  \end{subfigure}
  \hfill
  \begin{subfigure}{0.33\textwidth}
    \includegraphics[width=\textwidth]{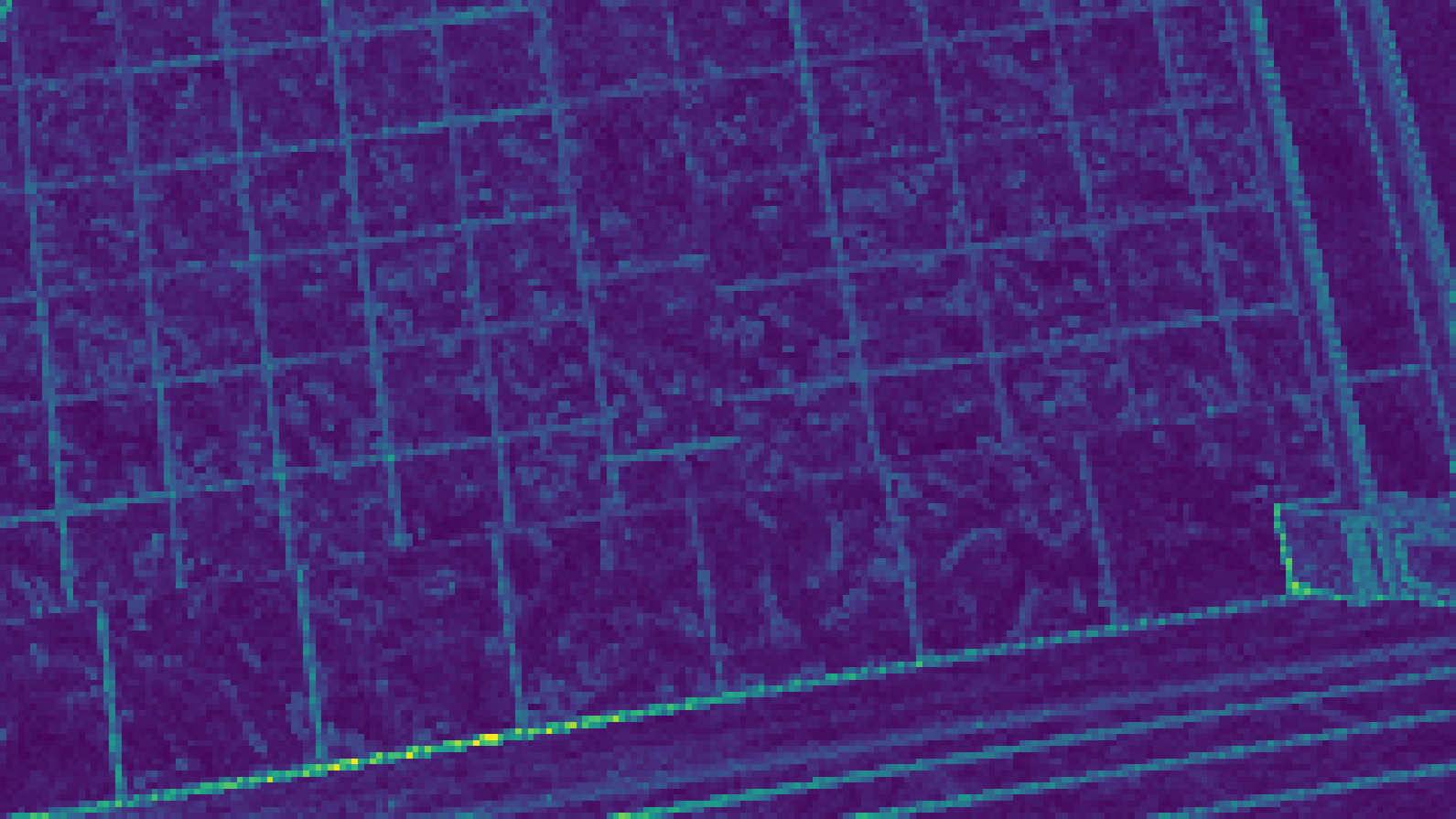}
    \caption{4x4}
  \end{subfigure}
  \vspace{.5em}
  \hfill
  \begin{subfigure}{0.33\textwidth}
    \includegraphics[width=\textwidth]{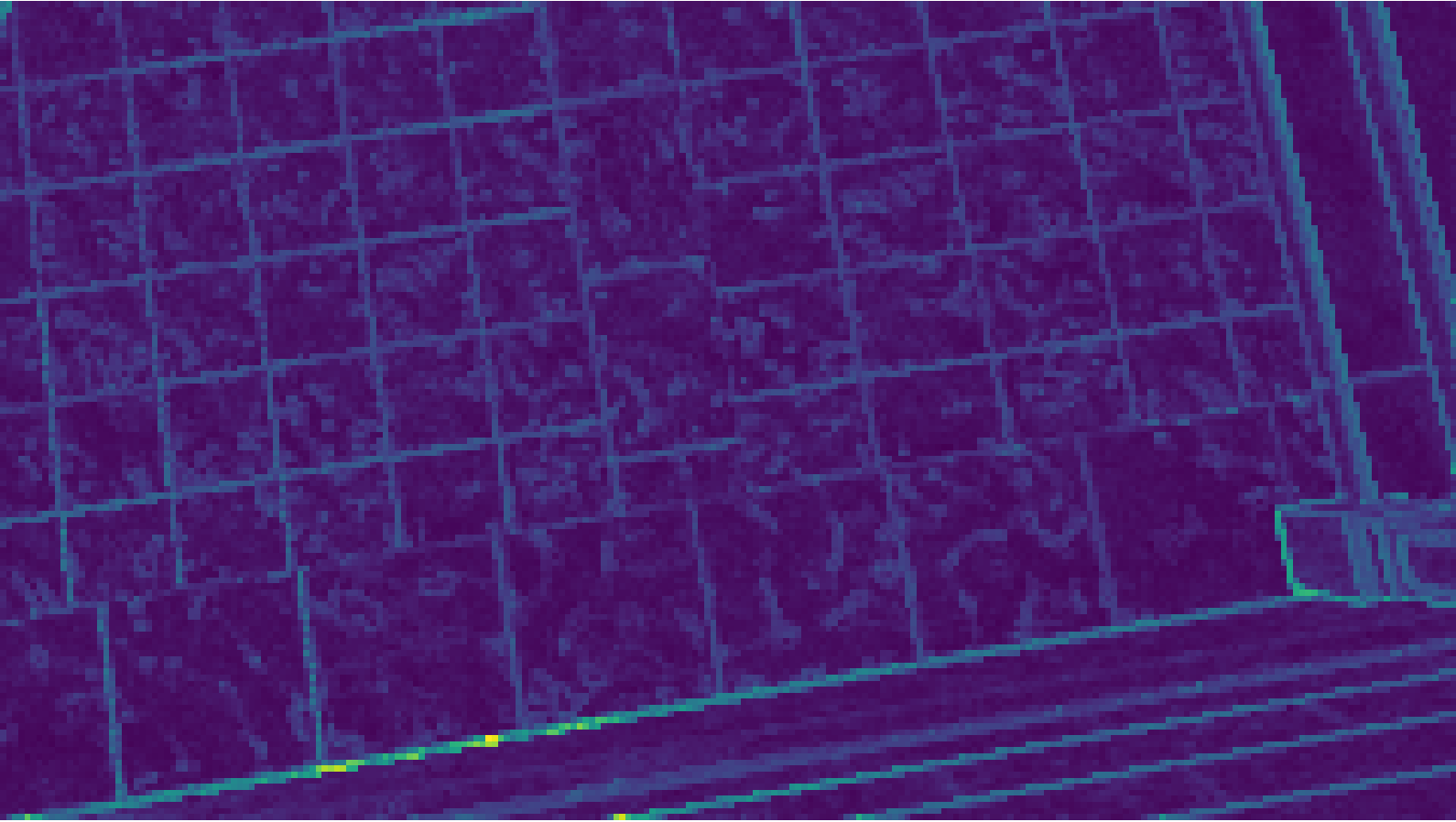}
    \caption{2x2 Extrapolated}
  \end{subfigure}
  \hfill
  \begin{subfigure}{0.33\textwidth}
    \includegraphics[width=\textwidth]{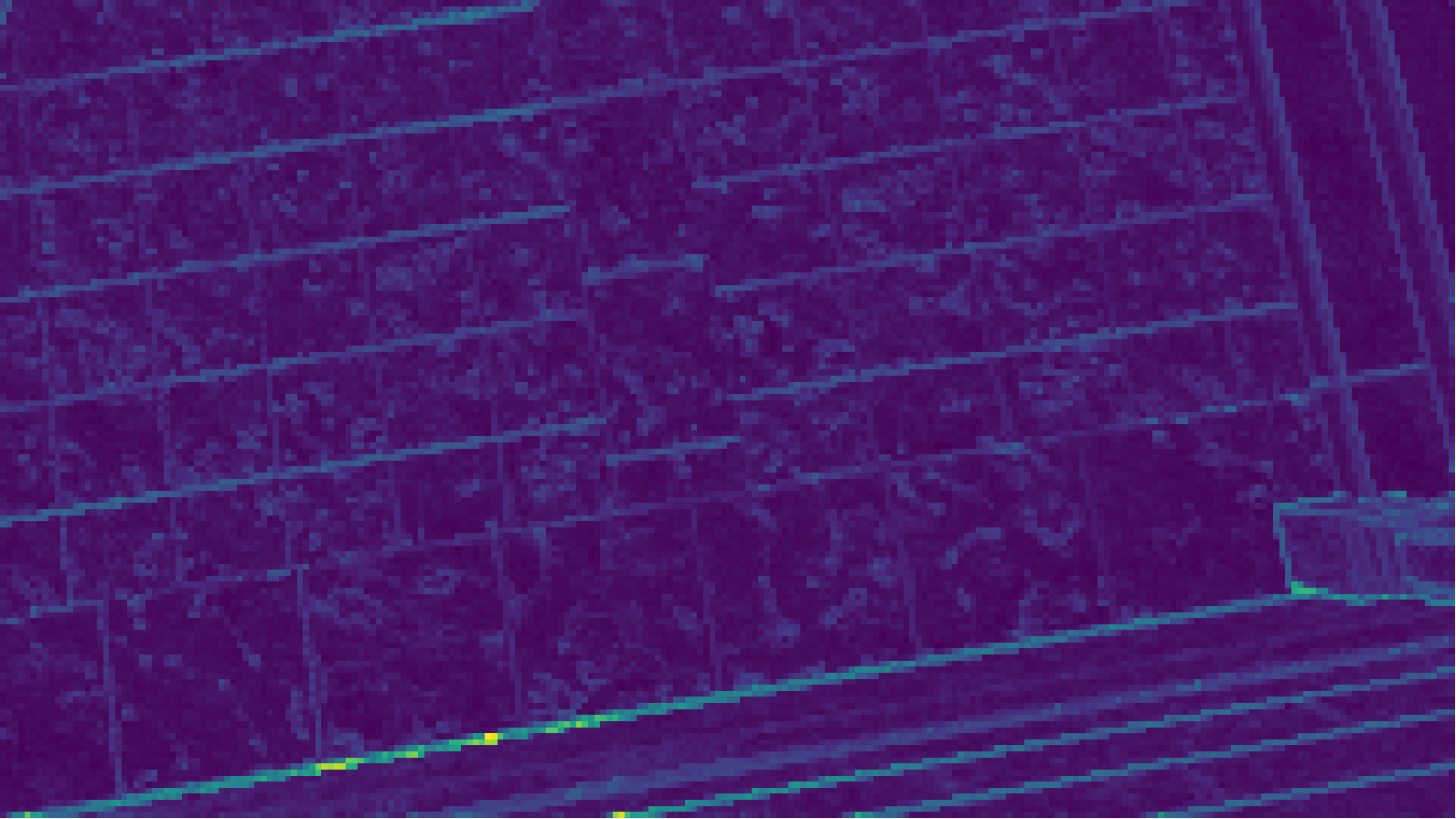}
    \caption{4x2 Extrapolated}
  \end{subfigure}
  \hfill
  \begin{subfigure}{0.33\textwidth}
    \includegraphics[width=\textwidth]{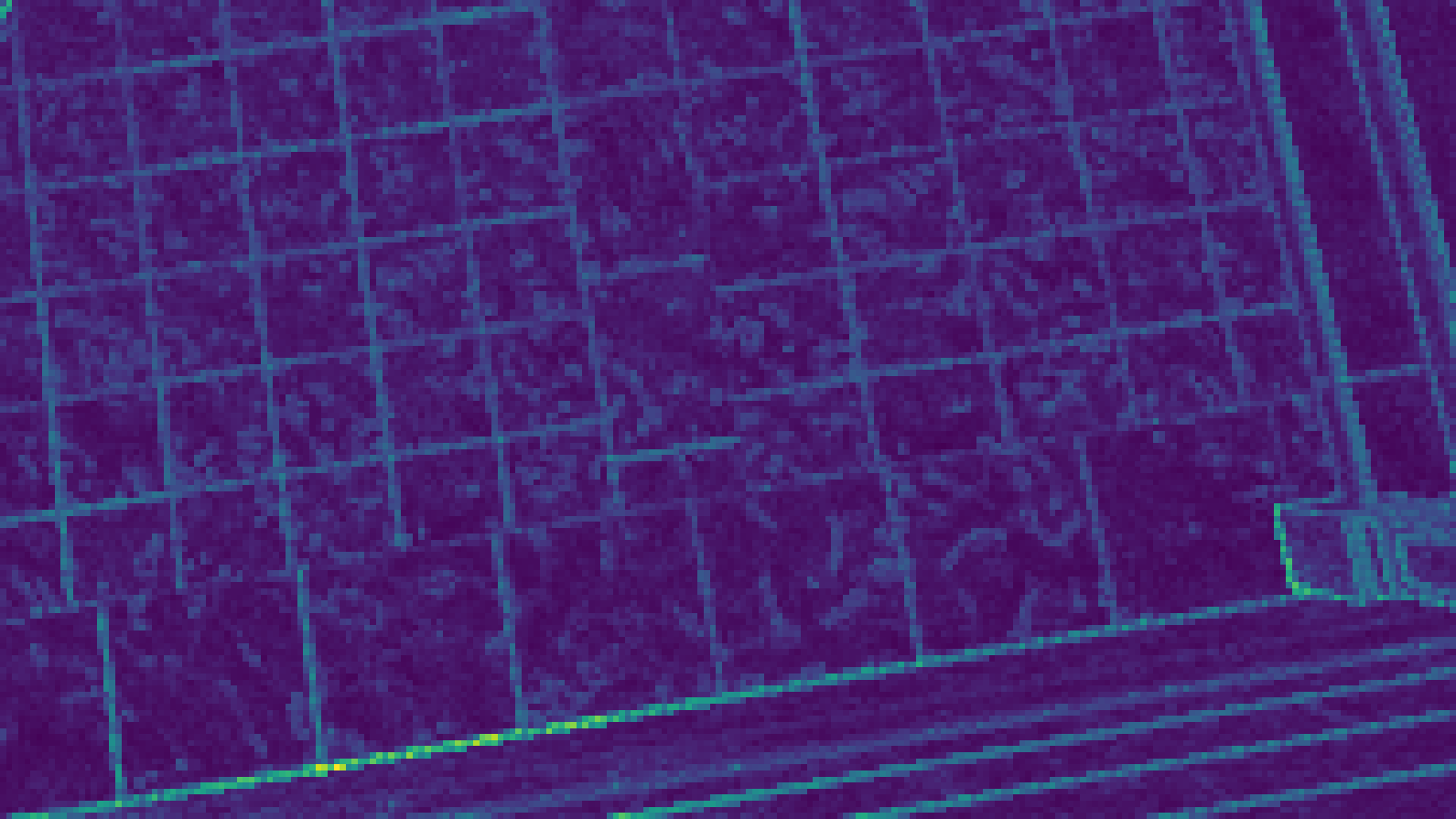}
    \caption{4x4 Extrapolated}
  \end{subfigure}
  \caption{FLIP at different shading rates. Ground truth versus extrapolation of them as described in Section \ref{sec:extrapolation}.}
\end{figure*}

\end{document}